\documentclass{pas}

\usepackage{multirow}
\RequirePackage{aas-macros}
\usepackage{orcidlink}
\usepackage{amsmath}
\usepackage[nopatch]{microtype}
\usepackage{booktabs}
\usepackage{listings}
\usepackage{graphicx}
\usepackage{graphbox} 
\usepackage{natbib}
\usepackage{hyperref}
\DeclareUnicodeCharacter{2212}{-}
\usepackage{lineno}
\usepackage{booktabs}
\usepackage{tabularx}
\usepackage{array}
\usepackage{yhmath}

\newcolumntype{C}{>{\centering\arraybackslash}X}

\begin{document}

\lefttitle{Publications of the Astronomical Society of Australia}
\righttitle{Van Kempen W, et al.}

\jnlPage{x}{xx}
\jnlDoiYr{2024}
\doival{xx.xxxx/pasa.xxxx.xx}

\articletitt{Research Paper}

\title{A Comprehensive Investigation of Environmental Influences on Galaxies in Group Environments}

\author{\sn{Wesley} \gn{Van Kempen}$^{1}$\orcidlink{0009-0009-8499-1326},
            \sn{Michelle E.} \gn{Cluver}$^{1,2}$\orcidlink{0000-0002-9871-6490},
            \sn{Thomas H.} \gn{Jarrett}$^{3,4,1}$\orcidlink{0000-0002-4939-734X},
            \sn{Darren J.} \gn{Croton}$^{1,5,6}$\orcidlink{0000-0002-5009-512X},
            \sn{Trystan S.} \gn{Lambert}$^{7,8}$\orcidlink{0000-0001-6263-0970},
            \sn{Virginia A.} \gn{Kilborn}$^{1,5}$\orcidlink{0000-0003-3636-4474},
            \sn{Edward N.} \gn{Taylor}$^{1}$\orcidlink{0000-0002-5522-9107}, 
            \sn{Christina} \gn{Magoulas}$^{9}$\orcidlink{0000-0001-9090-5548} and
            \sn{H. F. M.} \gn{Yao}$^{2,11}$\orcidlink{0000-0001-8459-4034}}

\affil{$^1$Centre for Astrophysics and Supercomputing, Swinburne University of Technology, John Street, Hawthorn, VIC 3122, Australia, $^2$Department of Physics and Astronomy, University of the Western Cape, Robert Sobukwe Road, Bellville, 7535, South Africa, $^3$Institute for Astronomy, University of Hawaii, 2680 Woodlawn Drive, Honolulu, HI 96822, USA, $^4$Department of Astronomy, University of Cape Town, Rondebosch, 7700, South Africa, $^5$ARC Centre of Excellence for All Sky Astrophysics in 3 Dimensions (ASTRO 3D), Melbourne, Australia, $^6$ARC Centre of Excellence for Dark Matter Particle Physics (CDM), Melbourne, Australia, $^7$Instituto de Estudios Astrofísicos, Facultad de Ingeniería y Ciencias, Universidad Diego Portales, Av. Ejército Libertador 441, Santiago, Chile, $^8$ICRAR, The University of Western Australia, 35 Stirling Highway, Crawley, WA 6009, Australia, $^{9}$Australian Synchrotron, 800 Blackburn Road, Clayton, VIC 3168, Australia and $^{10}$ Department of Physics, University of Félix Houphouët Boigny, 01 BP V 34 Abidjan 01, Côte d’Ivoire.}

\corresp{W. Van Kempen, Email: wvankempen@swin.edu.au}

\citeauth{Van Kempen W, Cluver M.E., Jarrett T.H., Croton D.J., Lambert T.S., Kilborn V.A., Taylor E.N., Magoulas C. and Yao H.F.M. (2024) A Comprehensive Investigation of Environmental Influences on Galaxies in Group Environments. {\it Publications of the Astronomical Society of Australia} {\bf 00}, x--xx. https://doi.org/xx.xxxx/pasa.xxxx.xx}

\history{(Received 23/07/2024; revised 08/10/2024; accepted 19/10/2024)}

\begin{abstract}

Environment has long been known to have an impact on the evolution of galaxies, but disentangling its impact from mass evolution requires the careful analysis of statistically significant samples. By implementing cutting-edge visualisation methods to test and validate group-finding algorithms, we utilise a mass-complete sample of galaxies to $z < 0.1$ comprised of spectroscopic redshifts from prominent surveys such as the 2-degree Field Galaxy Redshift Survey and the Galaxy and Mass Assembly Survey. Utilising our group finding methods, we find 1,413 galaxy groups made up of 8,990 galaxies corresponding to 36\% of galaxies associated with group environments. We also search for close pairs, with separations of $r_{\rm sep}<50$ $\text{h}^{-1}\text{kpc}$ and $v_{\rm sep}< 500 \: \text{km s}^{-1}$ within our sample and further classified them into major ($M_{sec}/M_{prim} \leq$ 0.25) and minor ($M_{sec}/M_{prim} >$ 0.25) pairs. To examine the impact of environmental factors, we employ bespoke WISE photometry, which facilitates accurate measurements of stellar mass and star formation rates and hence the best possible description of the variation of galaxy properties as a function of the local environment. Our analysis, employing a derived star-forming main sequence relation, reveals that star-formation (SF) within galaxies are pre-processed as a function of group membership. This is evident from the evolution of the star-forming and quenched population of galaxies. We see an increase in the fraction of quiescent galaxies relative to the field as group membership increases and this excess of quenched galaxies relative to the field is later quantified through the use of the \textit{environmental quenching efficiency} ($\epsilon_{env}$) metric. Within the star-forming population, we observe SF pre-processing with the relative difference in specific star formation rates ($\Delta sSFR$), where we see a net decrease in SF as group membership increases, particularly at larger stellar masses. We again quantify this change within the SF population with our \textit{star formation deficiency} ($\epsilon_{SFD}$) metric. Our sample of close pairs at low stellar masses exhibit enhanced star formation efficiencies compared to the field, and at larger stellar mass ranges show large deficiencies. Separating the close pairs into major/minors and primary/secondaries reveals SF enhancements projected separation decreases within the minor pairs, this effect is even more pronounced within minor primaries. This research emphasises the importance of carefully studying the properties of galaxies within group environments to better understand the pre-processing of SF within galaxies. Our results show that the small-scale environments of galaxies influence star-forming properties even when stellar masses are kept constant. This demonstrates that galaxies do not evolve in isolation over cosmic time but are shaped by a complex interaction between their internal dynamics and external influences.

\end{abstract}


\maketitle



\section{Introduction}
\label{sec:Introduction}

Since the results of early spectroscopic surveys \citep[e.g.][]{Kirshner81,deLapparent86} we have observed the distribution of galaxies in the universe to be inhomogeneous in nature with regions of high and low galaxy density. The term cosmic web describes this variance in galaxy density as it appears to form ``web'' like structures \citep{Bond96}. The cosmic web forms the large-scale structure (LSS) which is most prominently well-defined in the local ($z < 0.1$) universe, where filamentary networks connect regions of low density (voids) to high density (nodes) \citep[e.g.][]{Bond96, Colberg05,vandeWeygaert08, Arag´on-Calvo10}. The foundational theory governing the formation and evolution of the cosmic web was established within the framework of the Zel’dovich formalism \citep{Zel'dovich70}. This model predicts that cosmic web constituents arise due to the accretion of dark matter, driven by gravitational growth from small density and velocity perturbations during the early Universe, which is now well established within the $\Lambda$CDM paradigm \citep[e.g.][]{Davis85, Peebles03, Springel05}.

There is a clear connection between the characteristics of galaxies and the broader LSS in which they are embedded \citep[e.g.][]{Oemler74, Dressler80, Kauffmann04, Poggianti06, Boselli11, Taylor15, Bluck20}. In high-density environments, such as galaxy clusters, early-type galaxies dominate. These galaxies are characterised by their gas-poor, optical red colour, and passive star-forming nature. Conversely, in low-density environments, late-type galaxies are more prevalent. These galaxies are known for their gas-rich composition, optically blue colours, and ongoing star formation (SF). Typically, this relationship between morphology and density \citep[Morphology-Density relation;][]{Dressler97} is explained through various physical effects that occur in the exceedingly dense intra-cluster medium (ICM) that quench a galaxy's SF during its infall \citep{Rhee17, Finn23}. However, red passive galaxies can also be observed in lower-density environments such as galaxy groups \citep{Lietzen12, Vulcani15}. 

SF quenching, i.e. the process by which the formation of new stars in a galaxy is significantly reduced or stopped altogether, can take place before galaxies enter clusters, especially in lower-density environments like filaments and groups, a phenomenon commonly known as ``pre-processing'' \citep{Fujita04}. The concept of galaxies undergoing pre-processing before entering clusters has been explored through observational \citep[e.g.][]{Cortese06, Cluver2020, Estrada23, Lopes24} and simulation-based \citep[e.g.][]{Rhee17, Bakels21, Haggar23} studies. These studies demonstrate that indeed galaxies undergo pre-processing prior to cluster in-fall. However, when attempting to study the environmental effects on galaxies, disentangling and controlling for other critical processes that contribute to SF quenching, such as ``mass'' related quenching mechanisms \citep{Driver06, Peng10}, can pose a challenging task.

Mass quenching (or secular quenching) refers to the internal processes within a galaxy that results in the global quenching of SF \citep{Peng10}. The various mass quenching mechanisms, i.e. virial shock heating \citep{Birnboim03}, AGN feedback \citep[e.g.][]{Croton06}, the strangulation of cold gas \citep[e.g.][]{Larson80} and the presence of a bulge \citep[e.g.][]{Martig09}, are all more prevalent in galaxies with larger stellar mass.

On the other hand, environmental quenching includes all external processes related to a galaxy's surrounding environment that suppress star formation independently of the galaxy's stellar mass  \citep{Peng10}. Environmental quenching mechanisms can be sorted into two main categories. The first category encompasses gravitational perturbations, where actions such as the stripping or disturbance of a galaxy's gas due to frequent, high-velocity interactions with other galaxies \citep[galaxy harassment;][]{Moore96}, the removal of stellar and gas material from neighbouring galaxies \citep[tidal stripping;][]{Mihos04}, and the formation of a stellar spheroid or bulge resulting from galaxy-galaxy mergers \citep[morphological quenching;][]{Martig09}. The second category encompasses interactions within the intra-group or intra-cluster mediums i.e. a galaxy moving through the hot ($T_{ICM}\sim10^{7}-10^{8}$ K), highly dense ($n_{ICM}\sim10^{-4}-10^{-2}$ cm$^{-3}$) intra-cluster medium (ICM) at high velocities ($\sigma \approx 500-1000$ km s$^{-1}$) resulting in instabilities of the gas, that leads to the ionisation and removal of the gas \citep[ram pressure stripping;][]{Gunn72, Sarazin86, Boselli22}. Additionally, the differential viscous properties between a galaxy's interstellar medium (ISM) and the ICM or intra-group medium (IGM) result in frictional forces as the galaxy navigates the cluster or group environment. In the works of \cite{Walker10} and \cite{Cluver13}, compact groups have properties that suggest these frictional forces strip the star-forming gas from galaxies' interstellar medium, a phenomenon known as 'viscous stripping' \citep{Nulsen82}.

The impact of environmental quenching is closely correlated to the density of the surrounding environment. Consequently, its effects are more pronounced at lower redshifts and in regions of higher density \citep{Contini20, Einasto2020}. Given that approximately 40-50\% of all galaxies in the local universe $z=0$ universe are found within galaxy groups \citep{Eke04}, it's important that we can understand the various mechanisms and outcomes of these mechanisms on galaxy evolution.
This need is underlined by the works of \cite{Cluver2020}, which found evidence of pre-processing in the group environment by demonstrating that lower massed galaxies ($M_{\star} < 10^{10.5} M_{\odot}$) when in groups had an increased fraction of quenched galaxies when compared to non-group (``field'') galaxies with the same stellar mass. \cite{Lietzen12} congruently found that the fraction of SF galaxies declines and the fraction of quenched galaxies increases with an increase in group membership, and at a group membership of 8 members quenched galaxies become more prevalent than star-forming galaxies. \cite{Davies19} conducted an analysis using the Galaxy and Mass Assembly (GAMA) group catalogue established by \cite{Robotham11}, which demonstrated that the passive fraction of galaxies rises with increasing stellar and halo masses.

When considering close pair studies conducted by \cite{Lambas03} and \cite{Ellison10}, they independently demonstrated that environmental interactions within galaxy pairs commonly trigger SF as a function of projected separation. \cite{Woods10} found that major pairs had on average higher SF than minor pairs.

Utilising a large contiguous area of 384 square degrees using carefully vetted characterisations of the group environment in tandem with bespoke WISE photometry and derived properties, we have analysed a dataset that provides the statistical power to create a new benchmark view of environment for the $z < 0.1$ universe. This allows us to expand upon these previous studies by exploring the local environments of galaxies in close pairs and group environments and contrasting them to the isolated field setting, to understand the consequential effects of the key mechanisms that drive environmental quenching and SF pre-processing within galaxies. The effects of pre-processing within these environments can be observed through changes in SF when controlling for the effects of stellar mass, as galaxies evolve from late-type star-forming galaxies to early-type quenched systems \citep{Lietzen12, Barsanti18, Cluver2020}. This is particularly true for low-massed galaxies infalling to high-massed dark matter halos \citep{Roberts17, Boselli22}. 

The Wide-Field Infrared Survey Explorer \citep[WISE;][]{Wright2010} surveyed the whole sky in the mid-infrared (MIR) at the wavelengths of 3.4 $\mu$m (W1), 4.6 $\mu$m (W2), 12 $\mu$m (W3) and 23 $\mu$m \citep[W4;][]{Brown14}, this provides us with continuum emission descending from old stars provided from the W1 and W2 bands which trace dust-free stellar mass. The W3 and W4 bands trace dust reprocessed SF with the W3 band sensitive to the continuum emission as well as atomic and molecular lines and the W4 band sensitive to the small-grain dust continuum \citep{Jarrett13}. By utilising the power of WISE, it provides us with excellent fiducial measurements of stellar mass \citep{Jarrett23}, infrared-colours \citep{Jarrett13, Jarrett2019} and SF properties (Cluver et al. in prep) to isolate and quantify the effects of the environment.

Our objective is to examine the impacts of the local environment on diverse galaxy properties as inferred from WISE MIR observations. One of the most significant and challenging tasks to overcome when investigating environmental effects on galaxy evolution is to quantify and control for the local galaxy density. In this study we have created a group catalogue making use of the {\tt FoFpy} Python package by \cite{Lambert20} which utilises a graph-theory-based modified version of the friends-of-friends algorithm from \cite{Huchra82}. We have also produced a close-pair catalogue based on \cite{Robotham2014}. To produce robust, statistically developed groups, we fine-tuned the FoF algorithm using simulations produced with the {\tt Theoretical Astrophysical Observatory} (TAO) utilising the {\tt Millenium} dark matter simulation \citep{Springel05} and the Semi-Analytic Galaxy Evolution (SAGE) model \citep{Croton06}. For both the group and pair catalogues we followed up by visually inspecting these data sets with the 4D visualisation tool {\tt Partiview} \citep{Levy10}.

The area of focus for our study is centred on the Southern Galactic Pole (SGP) comprising 384 square degrees, and is encompassed within the optical imaging survey of the Kilo Degree Survey South \citep[KiDS-S;][]{dejong2013} region. This target field has also been the host of massive spectroscopic galaxy surveys, providing us with redshifts from both the 2-degree Field Galaxy Redshift Survey \citep[2dFGRS;][]{Colless01} and the G23 region of the Galaxy and Mass Assembly \citep[GAMA;][]{Driver09} survey. By making use of this area, we can cross-match the infrared photometry of WISE to that of the spectroscopic surveys.

This paper is organised as follows. In Section \ref{sec:Data and Sample Selection} we provide details of the data, sample, and selection parameters that form the foundations of the study. In Section \ref{sec:Analysis} we present our analysis, which includes the effects of local environment on star formation quenching and the star-forming population (Sections \ref{subsec:SFQ_Local} and \ref{subsec:SF_Local}), the effects of group environments on star formation quenching and the star-forming population (Sections \ref{subsec:SFQ_Group} and \ref{subsec:SF_Group}), the effects of close pair environments on star formation (Section \ref{subsec:Close Pair Effects on Star Formation}). We interpret and further quantify our analysis in our discussion in Section \ref{sec:Discussion}. Finally, we present a summary of our main results and general conclusions in Section \ref{sec:Sum_Con}.

Throughout this study, we have adopted the cosmological parameters of H$_{0}$ = 70 km s$^{-1}$ Mpc$^{-1}$, $\Omega_{M}$ = 0.3, and $\Omega_{\Lambda}$ = 0.7. We adopt a Chabrier (2003) initial mass function (IMF). All magnitudes are using the Vega magnitude system as discussed and utilised in the WISE photometric calibration in \cite{Jarrett11}.

\section{Data and Sample Selection}
\label{sec:Data and Sample Selection}

\begin{figure*}[!hbt]
\centering
    \begin{minipage}[t]{0.33\textwidth}
        \centering
        \includegraphics[width=\textwidth]{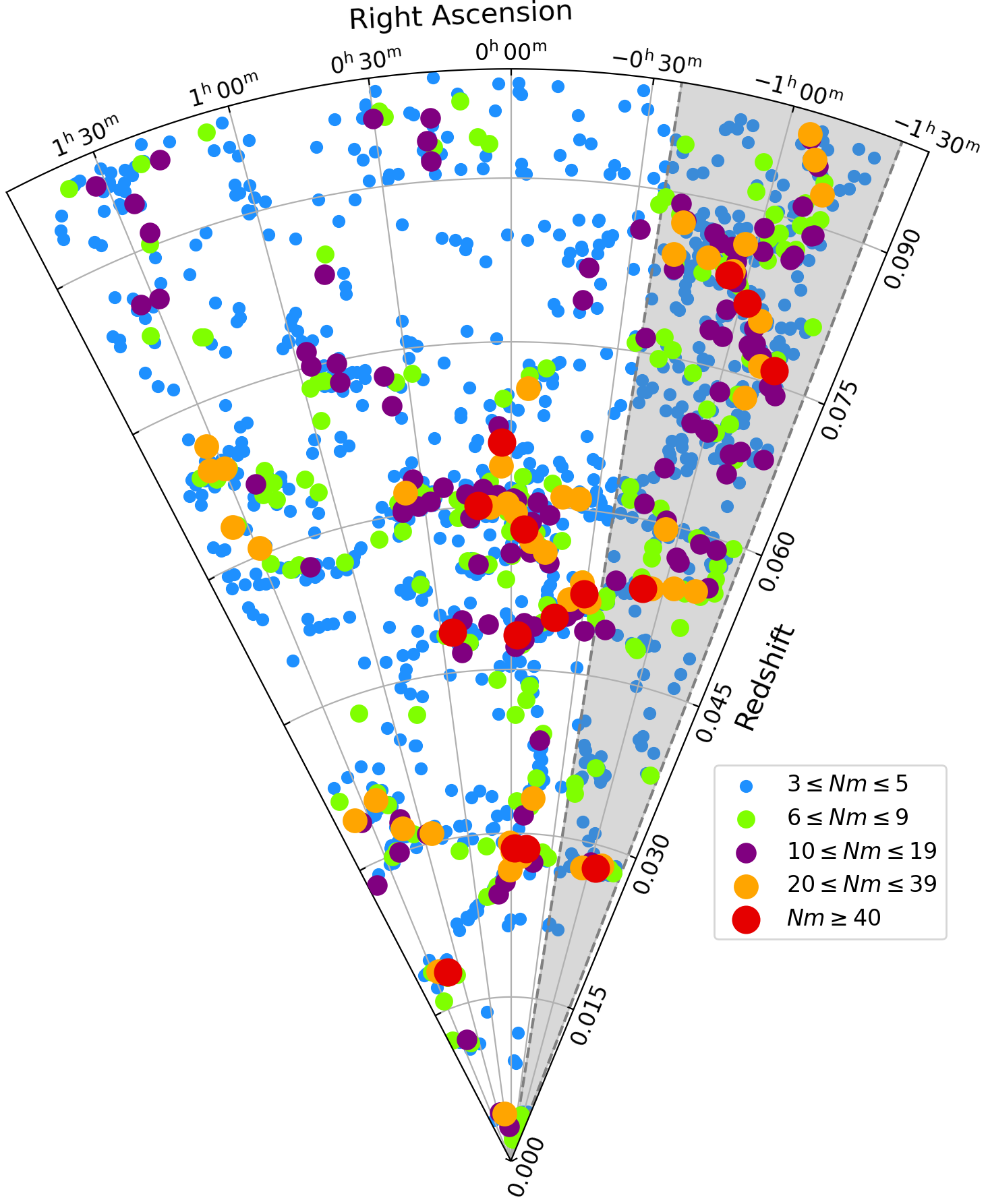}
        \parbox{\textwidth}{\centering \small (a) Galaxy Groups} \label{Groups}
    \end{minipage}\hfill
    \begin{minipage}[t]{0.33\textwidth}
        \centering
        \includegraphics[width=\textwidth]{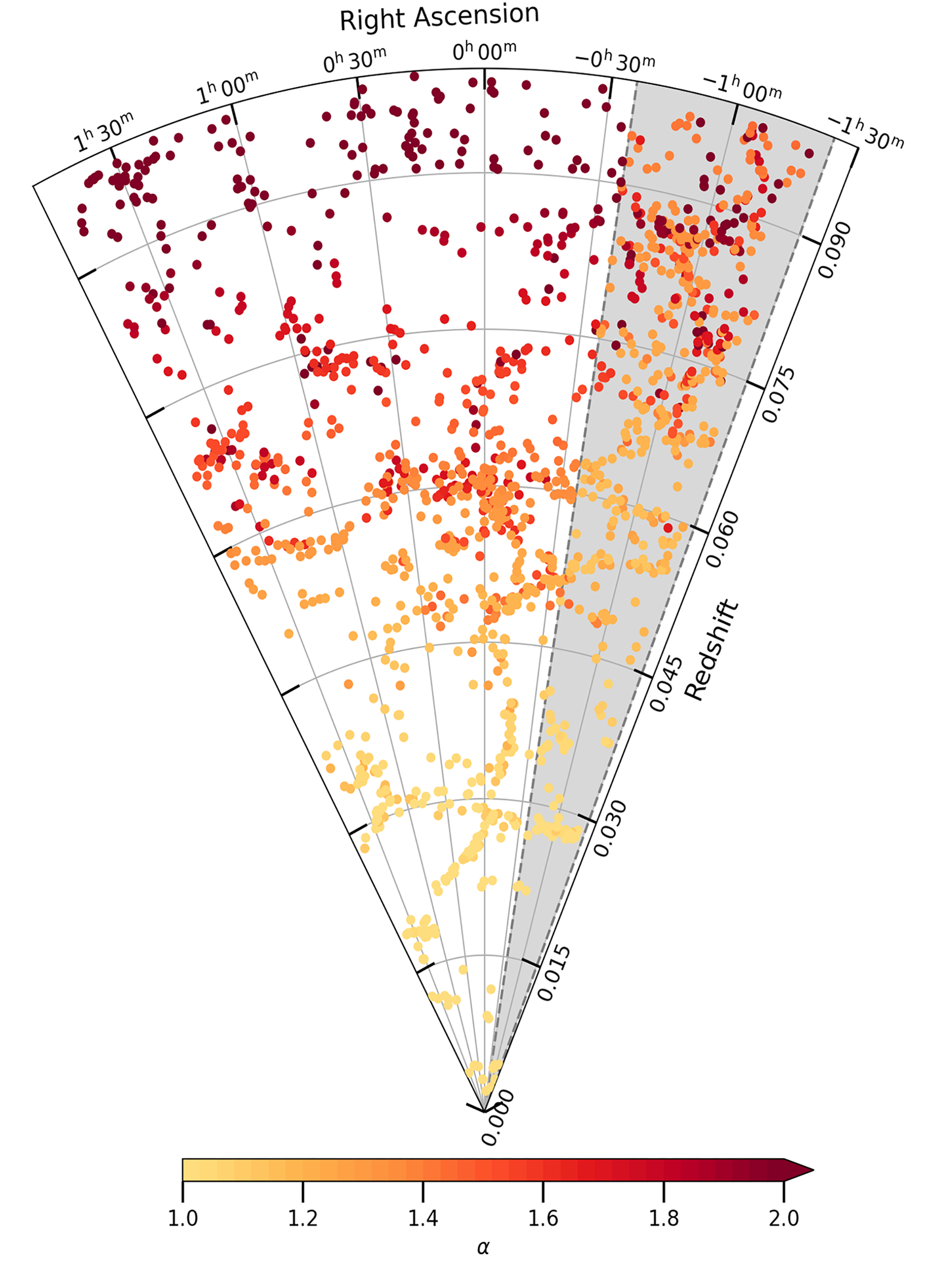}
        \parbox{\textwidth}{\centering \small (b) $\alpha$ Group Correction} \label{Groups Alpha}
    \end{minipage}\hfill
    \begin{minipage}[t]{0.33\textwidth}
        \centering
        \includegraphics[width=\textwidth]{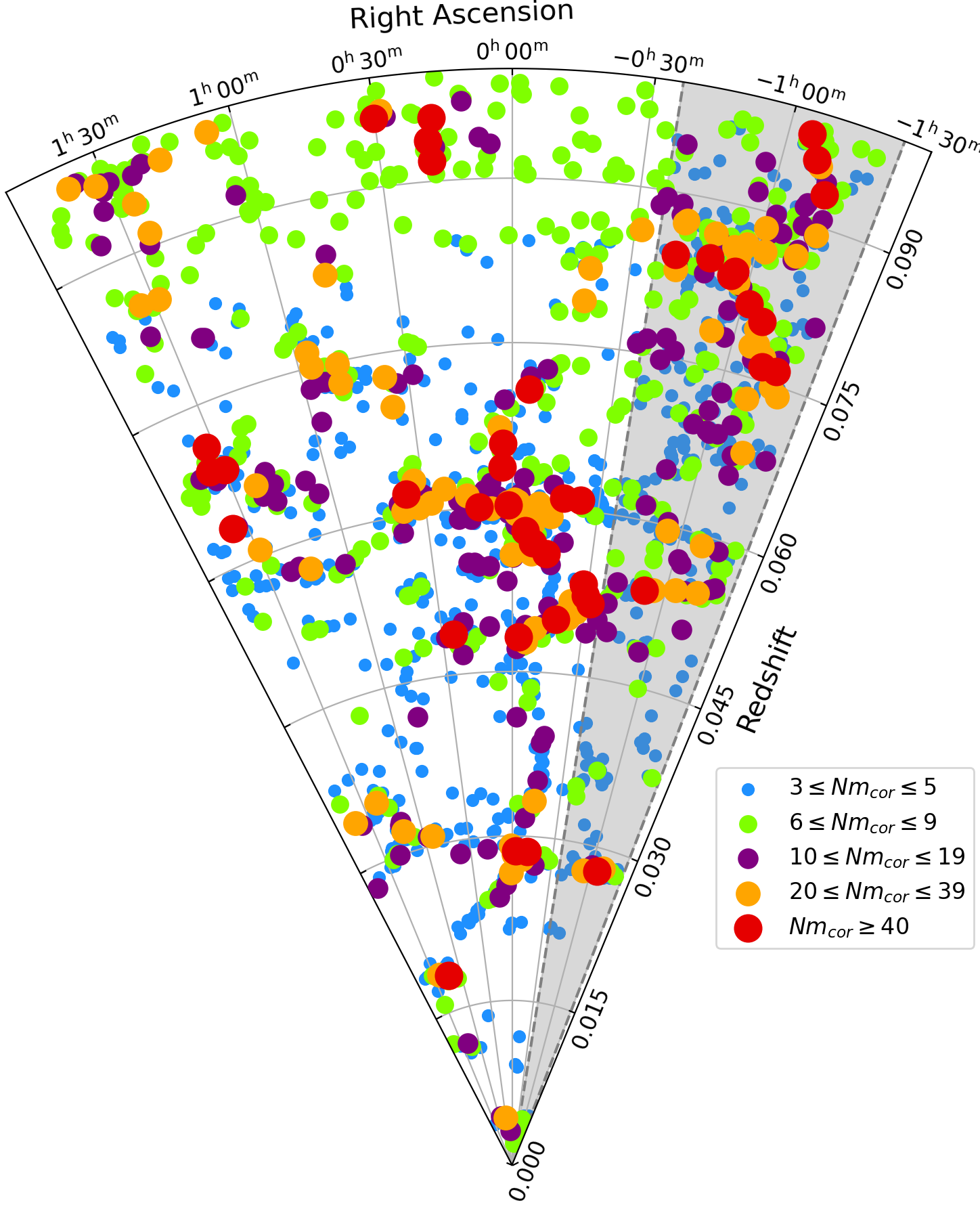}
        \parbox{\textwidth}{\centering \small (c) Corrected Galaxy Groups} \label{Groups_Cor}
    \end{minipage}
\caption{2D cone representation of the distribution of galaxy groups within the sample. Panel (a) illustrates galaxy groups binned by the number of members associated with each group by the FoF finder. Panel (b) depicts the $\alpha$ correction applied to each galaxy group, adjusting the total members within each group. Finally, panel (c) presents the distribution of galaxy groups binned by their corrected number of members after applying the $\alpha$ correction. The grey region indicates the GAMA G23 area within its right ascension limits.}
\label{Group_Completeness}
\end{figure*}

In this section, we outline the scope of our study and describe the meticulous construction of our sample. This includes the combination of data from various redshift surveys (Section \ref{subsec:SGP/KiDS-S Region}), resulting in contiguous sky coverage of 384 degrees squared. Section \ref{subsec:WISE Photometry and Derived Quantities} provides an overview of the WISE photometry, derived quantities, mass-complete and star-forming main sequence selections. We detail our approach to defining galaxy groups using the FoF algorithm in Section \ref{subsec:FoF Algorithm}, ensuring completeness control and justifying our choice of metric for the group environment in Section \ref{subsec:Group Completeness}. Our sample of close galaxy pairs is defined in Section \ref{subsec:Pair Selection}.

The redshift range is restricted to 0.1 to minimize systematic effects from observational biases and errors that grow with ($1+z$), and the systematic fall-off of mass sensitivity of the spectroscopic surveys and WISE W3 sensitivity (our key SF tracer).

\subsection{SGP Region}
\label{subsec:SGP/KiDS-S Region}


Our focal investigation is centred on the Southern Galactic Pole (SGP), covering an area of 384 square degrees within the celestial coordinates ranging from 340$^{\circ}$ to 26$^{\circ}$ in Right Ascension and -35.3$^{\circ}$ to -25.8$^{\circ}$ in Declination. The primary source of spectroscopic redshift measurements in our dataset is derived from the 2-degree Field Galaxy Redshift Survey (2dFGRS) and the Galaxy and Mass Assembly Survey \citep[GAMA;][]{Colless01,Driver09}. Within our dataset, 2dFGRS stands as the primary spectroscopic survey in our sample, having a magnitude limit of $b_{J} = 19.45$. However, in the G23 field (339.0$\leq$ RA $\leq$ 351.0, -35 $\leq$ Dec $\leq$ -30), the GAMA G23 survey surpasses the completeness of 2dFGRS by approximately 2.3 times, attributed to its fainter magnitude threshold of $i = 19.2$. Additionally, our dataset incorporates spectroscopic redshifts from the 2-degree Field Lensing Survey (2dFLenS), the 6-degree Field Galaxy Redshift Survey (6dFGRS), the 2MASS Redshift Survey (2MRS), and the Million Quasars catalogue (MILLIQUAS; \citep{Blake16,Jones04,Macri19,Flesch21}). However, these supplementary measurements constitute a minor portion of our study and represent only a small fraction of galaxies in their respective regions. This culminated effort results in a contiguous celestial coverage up to a redshift of 0.1, incorporating 24,656 spectroscopic sources. Further details regarding the survey origins for these sources are outlined in Table \ref{Surveys}. The 441 sources that are not accounted for in Table \ref{Surveys} originate from the WISE Extended Source Catalogue \citep[WXSC;][]{Jarrett13, Jarrett2019} and not any single spectroscopic survey\footnote{These sources are very large nearby galaxies and their redshift originates from the NASA/IPAC Extragalactic Database (NED) or Cosmicflows-3 catalog \cite{Tully16}.}.

\begin{table}[!hbt]
\centering
\caption{Sources of redshift measurements for the sample. 'Detected' indicates whether a given spectroscopic survey detected a source. It's worth noting that multiple surveys may detect the same source. The 'Used' column indicates the number of detected sources that served as the primary spectroscopic measurements in our sample.}
{\tablefont\begin{tabularx}{\columnwidth}{ C C C }  
\hline  \hline
Survey     &Detected      &Used  \\
\hline
2dFGRS           &19505     &17535   \\
GAMA             &3307      &3149   \\
2dFLenS          &2272      &2223   \\
6dFGRS           &2087      &835    \\
2MRS             &1643      &306    \\
MILLIQUAS        &208       &167    \\
\hline  \hline
\end{tabularx}}
\label{Surveys}
\end{table}

Ensuring a comprehensive understanding of the implications of combining these surveys into a contiguous region is crucial, as the resulting dataset exhibits inherent inhomogeneity attributable to the differing completeness levels of the individual surveys. This disparity necessitates a meticulous understanding and control of the dataset's inhomogeneity to ensure the reliability of subsequent analyses. Further details on the steps taken to address and mitigate this inhomogeneity are expanded upon in Section \ref{subsec:Group Completeness}.

\subsection{WISE Photometry and Derived Quantities}
\label{subsec:WISE Photometry and Derived Quantities}

\subsubsection{WISE Photometry}
\label{subsubsec:WISE Photometry}

All resolved galaxies are identified and measured as part of the WISE Extended Source Catalogue \citep[WXSC;][]{Jarrett13,Jarrett2019}, whereas the compact sources utilise photometry from the ALLWISE, point-source catalogue with appropriate aperture corrections to capture the total flux, including extended emission beyond the PSF \citep{Cutri12}. The WISE source characterisation and photometry extraction for the WXSC and ALLWISE catalogues are detailed in \cite{Cluver14,Cluver2020} and later demonstrated that the two catalogues can be interconnected so they may be utilised in tandem on the same system in the work of \cite{Jarrett23}. The spectroscopic sources have been systematically cross-referenced with both the WXSC and ALLWISE catalogues. Among the 24,656 spectroscopic sources under consideration, 22,933 were successfully cross-matched, yielding a completeness rate of 93\%.
\subsubsection{WISE Stellar Masses}
\label{subsubsec:WISE Stellar Masses}

An important aspect of studying galaxy evolution is the reliance on accurate estimations of
the fundamental physical properties of galaxies. All stellar mass ($M_{\star}$) measurements are derived as outlined in \cite{Jarrett23}, which improves upon the previously established $M_{\star}$-MIR relation in \cite{Cluver14}. The new stellar mass relations are formulated through the use of three weighted mass-to-light (M/L, $\Upsilon_{\star}$) scaling relations. Through the combined use of these relations, a galaxy's stellar mass can be derived within 0.10-0.12 dex accuracy for galaxies with W1 luminosity $> 10^{9}$ L$_{\odot}$ and 0.15-0.25 dex accuracy for galaxies with $< 10^{9}$ L$_{\odot}$.

The first of these relations incorporates the WISE W1 band 3.4 $\mu$m flux to produce a third-order cubic polynomial M/L relation ($\Upsilon_{\star}^{3.4\mu m}$):

\begin{equation}
\centering
\label{SM1}
\begin{split}
\log M_{\star} = A_{0} + A_{1} \cdot & \log L_{W1} + A_{2} \cdot (\log L_{W1})^2 \\
&+ A_{3} \cdot (\log L_{W1})^3,
\end{split}
\end{equation}

\noindent where the A coefficients are -12.62, 5.0, -0.44 and 0.016 respectively, and $L_{W1}$ is the W1 band rest-frame luminosity in units of L$_{\odot}$.

 The W1 M/L is a good initial measurement of a galaxy's stellar mass but does not handle some of the varying physical properties of galaxies. In particular, late-type galaxies have young stellar populations with large variances in morphology types. The W1 M/L relation is a good approximation when dealing with early-type galaxies with old stellar populations. To enhance the W1 M/L relation and account for such variations within galaxies, photometric colours can be incorporated to better indicate the dynamical features of a galaxy. Thus, when WISE colour information is available, the W1 M/L scaling relation is downgraded by a factor of 5 in the weighted stellar mass calculation. The WISE colours adopted by \cite{Jarrett23} are the W1-W2 and the W1-W3 colours with their respective equations:

\begin{equation}
\label{SM2}
\log \Upsilon_{\star}^{3.4\mu m} = A_{0} + A_{1} \cdot (C12),
\end{equation}

\noindent where the A coefficients are -0.38 and -1.05 respectively and C12 is the WISE W1-W2 colour in Vega magnitudes.

\begin{equation}
\label{SM3}
\log \Upsilon_{\star}^{3.4\mu m} = A_{0} \cdot \exp(-10^{0.4(C13-A_{1})}),
\end{equation}

\noindent where the A coefficients are 0.45 and 4.69 respectively and C13 is the WISE W1-W3 colour in Vega magnitudes.

The combined weighted mean of the calculated stellar mass from the three M/L relations of Equations \ref{SM1}, \ref{SM2} and \ref{SM3} have relatively low scatter. Due to the availability of well-derived WISE measurements in the SGP region, 98\% (22,559) of our WISE sources have reliable W1-W2 colours and 65\% (14,884) have reliable W1-W3 colours. We therefore expect high-quality derived stellar mass measurements from the M/L relation from \cite{Jarrett23}.

\subsubsection{WISE Mass Complete Sample}
\label{subsubsec:Mass Comp}

\begin{figure}[!hbt]
\centering
\includegraphics[width=0.98\linewidth]{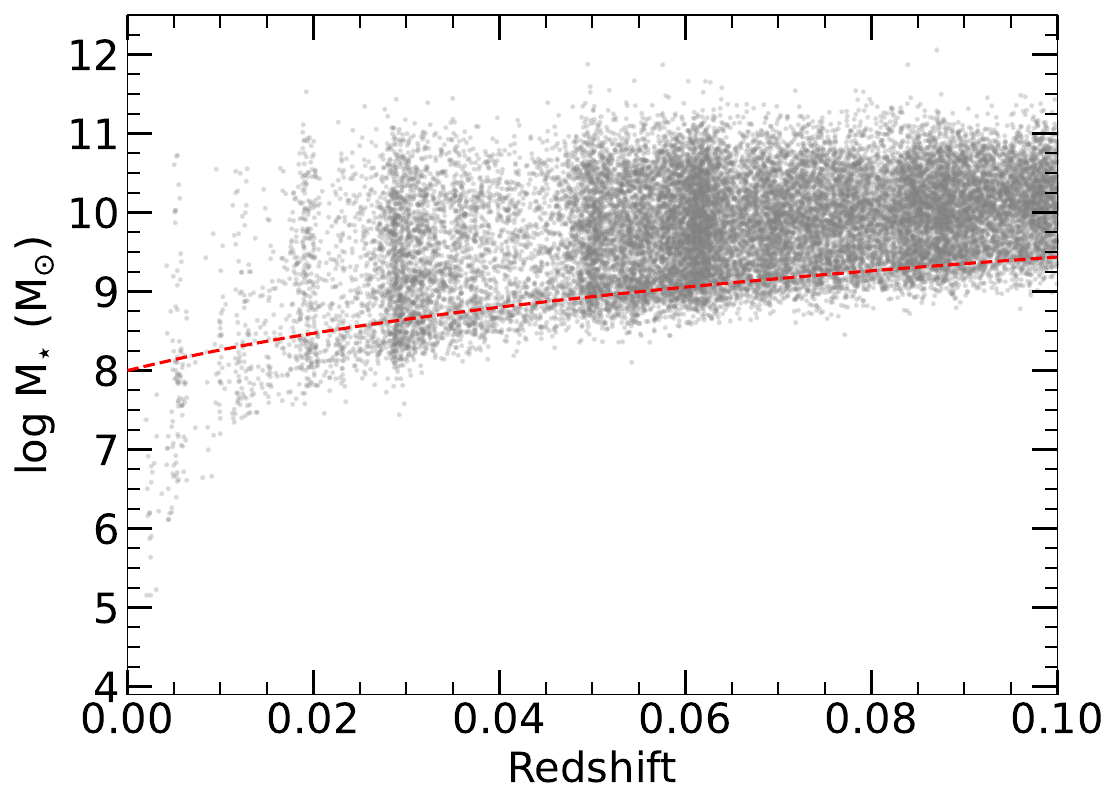}
\caption{WISE derived stellar masses as a function of redshift. The red dashed line indicates our redshift-dependent stellar mass completeness cut for our sample.}
\label{SM_cut}
\end{figure}

In our primary analysis, which focuses on stellar mass relationships, it is essential to use a mass-complete sample to ensure robust and unbiased results. To achieve this, we define our mass completeness over the redshift range of interest ($z<0.1$), implementing a carefully redshift-dependent calibrated stellar mass cut that follows the GAMA mass completeness methodology from \cite{Taylor15}. This selection enhances the statistical power of our analysis, particularly by increasing the number of low-mass galaxies, a key area of interest. Additionally, given the minimal evolutionary changes within this redshift range, this cut ensures that our sample remains representative whilst minimising potential biases. Figure \ref{SM_cut} visually demonstrates the redshift-dependent mass completeness applied in our analysis. Thus, over our redshift range, our stellar mass completeness is in the form:

\begin{equation}
\centering
\label{Completeness_Equation}
\log M_{\star} \: (M_{\odot}) = A_0 \cdot \frac{\ln(A_1 \cdot z +A_2)}{\ln A_3} + A_4,
\end{equation}

\noindent where the A coefficients are 0.45, 26.5, 1, 1.5 and 8 respectively. This redshift dependent stellar mass cut is applied throughout the entire analysis unless, explicitly stated otherwise (Section \ref{subsec:Close Pair Effects on Star Formation}).

\subsubsection{WISE Star Formation Rates}
\label{subsubsec:WISE Star Formation Rates}

Building on the work of \cite{Cluver2020} we measure star formation rates (SFR) using only WISE which provides a consistent set of measurements across a wide area and is well-matched in sensitivity for star-forming galaxies to $z<0.1$. This is additionally an opportunity to test the performance of these SFRs across different environment regimes. SFRs are derived using the relations in Cluver et al. (under review) which updates the calibration of \cite{Cluver2017} based on the relationship between WISE W3 and W4 to total infrared luminosity (LTIR). We include the SFR$_{W3}$ and SFR$_{W4}$ equations here for convenience:

\begin{equation}
\centering
\label{SFR_12}
\begin{split}
\log SFR_{W3} \: (M_{\odot} \: yr^{-1}) = 0.89  (\pm0.02) \log L_{W3} (L_{\odot}) \\
- 7.93 (\pm 0.20),
\end{split}
\end{equation}

\begin{equation}
\centering
\label{SFR_23_1}
\begin{split}
\log SFR_{W4} \: (M_{\odot} \: yr^{-1}) = 0.91  (\pm0.03) \log L_{W4} (L_{\odot}) \\
- 8.02 (\pm 0.23),
\end{split}
\end{equation}

\begin{equation}
\centering
\label{SFR_23_2}
\begin{split}
\log SFR_{W4} \: (M_{\odot} \: yr^{-1}) = 0.89  (\pm0.02) \log L_{W4} (L_{\odot}) \\
- 8.18 (\pm 0.19),
\end{split}
\end{equation}

\noindent where for sources with $\log (L_{W3}/L_{W4}) \geq -0.1$, Equation \ref{SFR_23_1} is used to calculate the SFR$_{W4}$ and for sources with $\log (L_{W3}/L_{W4}) < -0.1$, Equation \ref{SFR_23_2} is used.

SFRs are determined using an invariance-weighting of SFR$_{W3}$ and SFR$_{W4}$ and include a correction to account for star formation in low-mass galaxies with relatively low dust content, as determined in the nearby universe using the relationship between global dust density and UV-to-IR emission.
\subsubsection{WISE Colour-Colour Relation}
\label{subsubsec:WISE Colours}

\begin{figure}[!hbt]
\centering
\includegraphics[width=\linewidth]{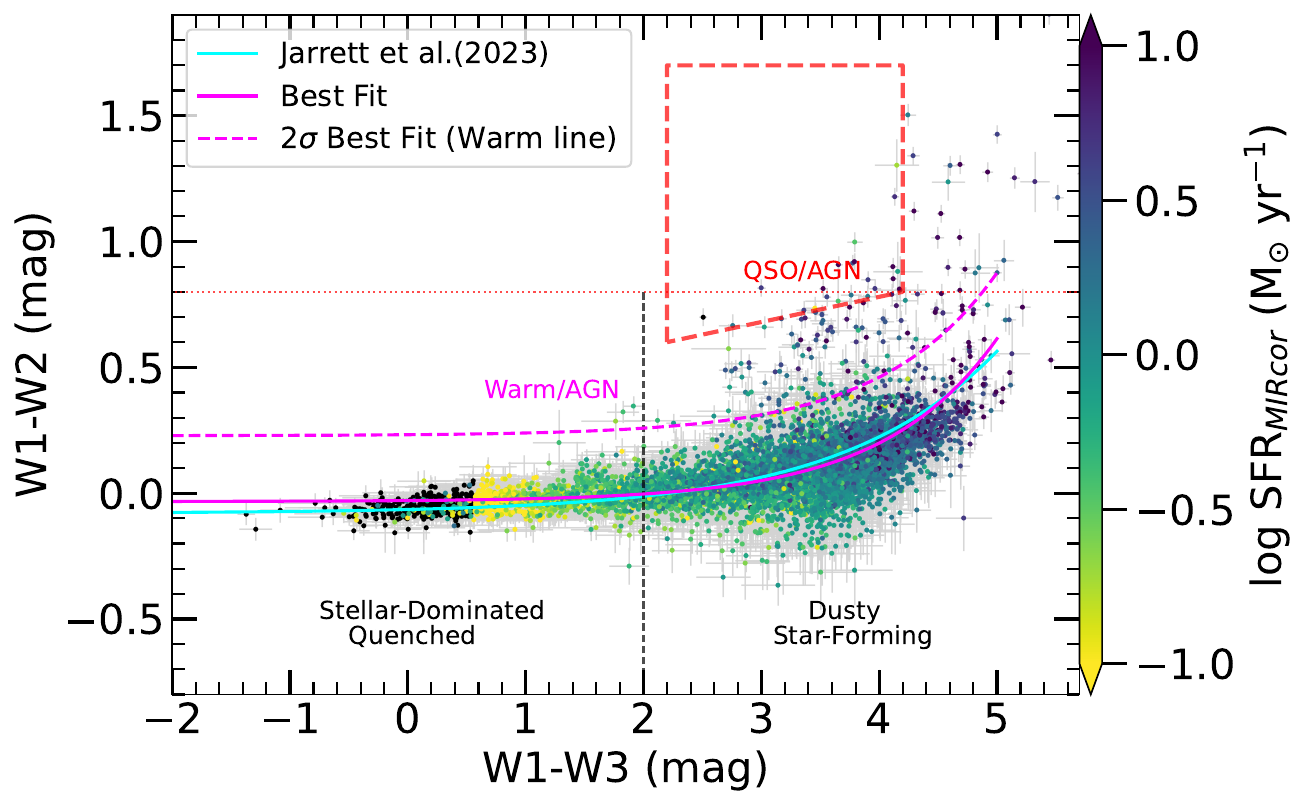}
\caption{WISE mid-infrared colour-colour (W1-W2 vs. W1-W3) distribution of the sample. The points are colour-coded by SFR, where the black points represent SFR UL. The black dashed line delineates stellar-dominated, quenched galaxies (W1-W3 $<$ 2) and dusty star-forming galaxies (W1-W3 $>$ 2). The red dotted line indicates the AGN threshold from \cite{Stern12} and the red dashed lines designate the QSO/AGN region from \cite{Jarrett11}. The cyan fit indicates the colour-colour sequence from \cite{Jarrett23}. The magenta and dashed magenta lines are fit to the distribution of points and the upper 2$\sigma$ offset, which indicates mid-infrared ``warm'' galaxies.}
\label{W1W2_W1W3}
\end{figure}

We produce the WISE colour-colour relations of WISE W1-W2 versus W1-W3 for the mass-complete sample to help refine our star-forming main sequence (SFMS) relation in Section \ref{subsubsec:Star-Forming Main Sequence}. In Figure \ref{W1W2_W1W3}, we present the mid-IR colour-colour (W1-W2 and W1-W3) relation for galaxies with reliable measurements as indicated by the signal-to-noise (S/N; S/N $>$ 7 and $>$ 3, respectively), excluding colour-colour upper limits (UL). The galaxies are colour-coded based on their corrected Mid-Infrared (MIR$_{cor}$) Star Formation Rate (Cluver et al. under review). We incorporate the designated regions outlined by \cite{Jarrett2019}, along with the colour-colour sequence fit from \cite{Jarrett23}.

Figure \ref{W1W2_W1W3} serves as a valuable diagnostic tool for discerning distinct galaxy populations, specifically, those dominated by stellar processes (quenched), engaged in dusty SF, and hosting AGN systems \citep{Yao20}. As anticipated from the colour-colour relation, a clear trend emerges, with SF generally being lowest at small W1-W3 colour values and increasing towards the right. Conversely, the disjunction in W1-W2 colour between the primary galaxy population and those at larger or redder W1-W2 values effectively segregates the main galaxy population from extreme cases and AGN. This demarcation is particularly advantageous, considering that sources falling into these categories are prone to yielding unreliable MIR-derived stellar masses and SFRs due to potential contamination from AGN hot-dust accretion \citep[see][]{Jarrett11, Stern12}.

In addressing these challenges, we follow the works of \cite{Yao22} and establish a conservative warm/AGN delineation by employing the best fit and augmenting the W1-W2 offset by 2$\sigma$ of W1-W2. This delineation effectively segregates extreme cases and AGN from the primary galaxy population, which will be utilised to help constrain the star-forming main sequence in Section \ref{subsubsec:Star-Forming Main Sequence}. The WISE colour-colour best fit and offset are expressed by the following equation:

\begin{equation}
\label{W1W2_W1W3_Equation}
W1-W2 = A_0 \cdot \exp\left(\frac{W1-W3}{A_1}\right) + A_2 + 2\sigma.
\end{equation}

\noindent Here, the best-fit coefficients A are determined to be 0.004, 1.01, and -0.029, respectively, with a standard deviation of 0.26. The fit to the WISE colour-colour relation exhibits high precision and closely aligns with the fit reported by \cite{Jarrett23}, where a substantially higher S/N was employed in establishing the relation.

\subsubsection{WISE Star-Forming Main Sequence}
\label{subsubsec:Star-Forming Main Sequence}

To investigate the environmental impact on a galaxy's star formation rate, it is crucial to understand the baseline relationship between star formation and stellar mass (the star formation history of the galaxy) \citep[e.g.,][]{Vulcani15, Trussler20, Finn23, Lopes23}. The SFR-stellar mass relation serves as a potent intrinsic tool, facilitating the classification of galaxies into either efficiently or inefficiently star-forming \citep{Peng10, Vulcani15, Calvi18, Bluck20, Epinat23, Goubert24}. By scrutinising this relationship across diverse local environments, such as galaxy pairs or groups, we can discern how environmental factors shape the SF properties of galaxies.

\begin{figure}[!hbt]
\centering
\includegraphics[width=\linewidth]{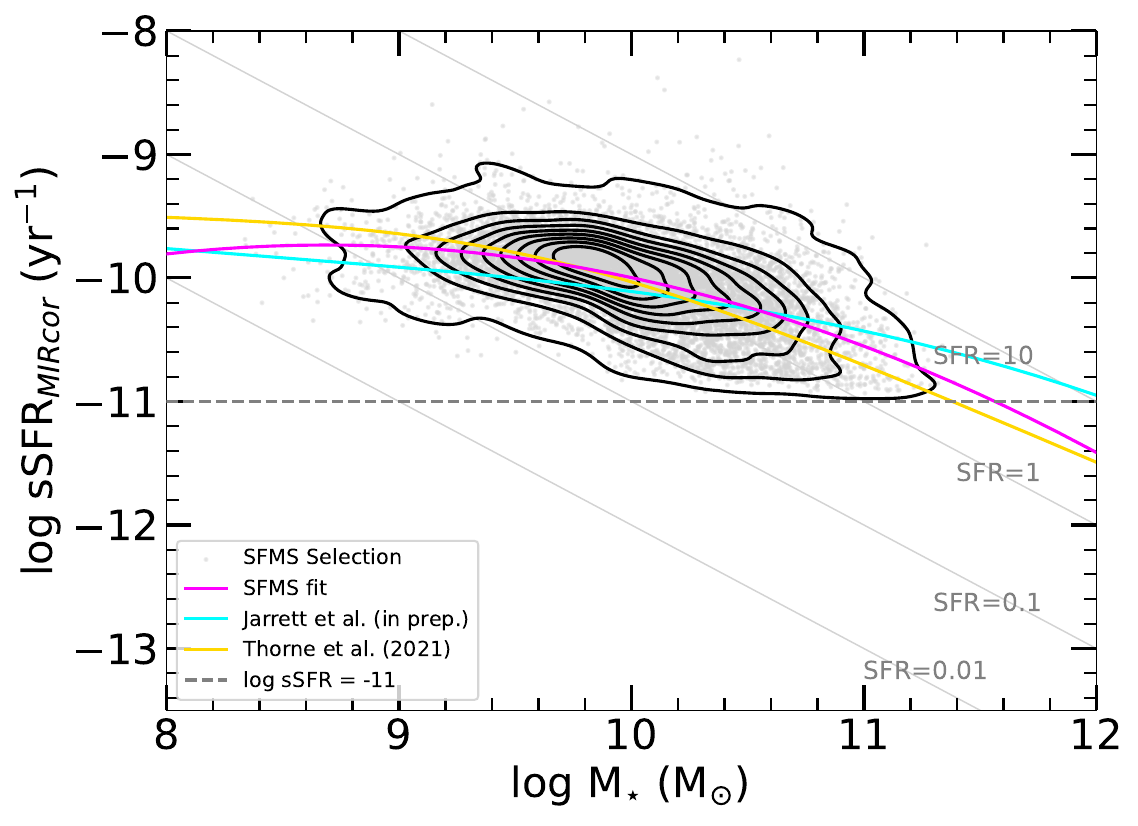}
\caption{The figure illustrates the star-forming main sequence (SFMS) illustrated with the sSFR vs. stellar mass relation. Grey points denote SFMS galaxies, whilst the black contours display the density of their distributions. The magenta solid line represents the 2nd-order polynomial fit for the SFMS. The grey dashed line indicates the $\log sSFR = -11.0$ threshold for segregating star-forming and quenched galaxies and the light grey solid lines denote lines of constant star formation.}
\label{SFMS}
\end{figure}

Central to examining this relationship is the concept of the SFMS, which refers to the region where efficient star-forming galaxies reside. This relationship can be differentiated through the SFR or specific star formation rate (sSFR; $SFR/M_{\star}$) - stellar mass relation. It provides us with the means to extract the typical rate at which galaxies of varying masses convert gas into stars, providing a benchmark for the typical SF behaviour.

To fit the SFMS, we utilised our redshift-dependent mass complete sample (see Section \ref{subsubsec:Mass Comp}) and removed quiescent galaxies by applying a $\log sSFR \geq -11.0$ threshold, a commonly used SF quenching separator \citep[e.g.][]{Houston23, Oxland24} and a W1-W3 $\geq 2$ colour cut \citep{Jarrett23} to remove stellar dominated galaxies and keep only dusty, star-forming galaxies. Additionally, to ensure the accuracy and reliability of our SFMS relation, we implement several quality criteria: the first involves removing SFR UL and low SFR S/N ratios (S/N $<$ 2), while the second entails excluding galaxies with a W1-W2 colour exceeding the warm line defined by Equation \ref{W1W2_W1W3_Equation}, as illustrated in Figure \ref{W1W2_W1W3}. These quality checks safeguard the SFMS against contamination from infrared-warm galaxies (e.g. AGNs) and low-quality measurements. The galaxies removed by the SFMS selection cuts can be viewed in Figure \ref{SFMS_Appendix} in Appendix \ref{sec:Appendix SFMS}. By employing these methodologies, we establish a robust SFMS, depicted in Figure \ref{SFMS}, with a best-fit relation expressed by the equation:

\begin{equation}
\centering
\label{SFMS_Equation}
\begin{split}
\log SFR \: (M_{\odot} \: yr^{-1}) = A_0 \log M_{\star}^{2} \: (M_{\odot}) \\
+ A_1 \log M_{\star} \: (M_{\odot}) + A_2,
\end{split}
\end{equation}

\noindent the coefficients $A_0$, $A_1$ and $A_2$ are determined to be -0.15, 3.65, and -21.22, respectively. In the next Sections we make use of our derived quantities to explore the properties and behaviours of the galaxies in our data set as a function of environment.

\subsection{Group Finding}
\label{subsec:FoF Algorithm}

To identify and establish galaxy groups, we have adopted a friends-of-friends (FoF) approach using the {\tt FoFpy} \footnote{The {\tt FoFpy} package, a Python implementation of the Friends-of-Friends (FoF) algorithm, is available at: \url{https://github.com/TrystanScottLambert/pyFoF}}. Python package from \cite{Lambert20}. Based on graph theory principles, the friends-of-friends algorithm aims to establish associations between galaxies, determining whether they are gravitationally linked based on their proximity in both projected and radial velocity space. The algorithm treats galaxies as nodes in a graph and utilises links to establish connections between these nodes.

The {\tt FoFpy} software package provides a range of inputs. The first pair of inputs are $D_{\rm initial}$ and $D_{\rm final}$. These represent the scaling factors used by the algorithm to determine the angular separation limits (see $D_{0}$ in Eq. 1 in \cite{Lambert20} for an explicit formalism). The projected angular separation of two galaxies is calculated as:

\begin{equation}
\label{FOF1}
D = \sin \left(\frac{\theta_{ij}}{2} \right) \cdot \frac{v_{\text{avg}}}{H_{0}},
\end{equation}

\noindent where $\theta_{ij}$ is the angular separation of two galaxies and $v_{avg}$ is the average line-of-sight velocity of two galaxies, where the line-of-sight velocity is $cz$. The non-relativistic velocity formula is adopted as we are constrained to the nearby universe.

Additionally, the input parameters $v_{\rm initial}$ and $v_{\rm final}$ signify the minimum and maximum disparities in line-of-sight velocities between the galaxies, calculated as:

\begin{equation}
\label{FOF2}
v = |v_{i} - v_{j}|.
\end{equation}

Moreover, the algorithm integrates rigid constraint parameters, denoted as $D_{\rm max}$ and $v_{\rm max}$, specifying the maximum allowable angular and line-of-sight velocity separations for the entire group. The number of trials ($n_{\rm trials}$) specifies the number of times the algorithm will run. During each trial, the algorithm integrates between various steps of the initial and final parameters, considering different starting points depending on the number of trials specified. Lastly, the \textit{Cutoff} parameter determines the probability threshold at which the algorithm stops considering two nodes as linked. When two nodes are associated through a link, the algorithm assigns a probability value between 0 and 1 to express the likelihood of this linkage, and the cutoff parameter sets the threshold below which the association is disregarded.

To enhance the output of the FoF algorithm, we incorporated simulations conducted through TAO, utilising the {\tt Millennium} dark matter simulation \citep{Springel05} and the Semi-Analytic Galaxy Evolution model \citep[SAGE;][]{Croton06}, and optimised the inputs for the FoF finder. We began by applying a mass cut of $M_{\star} > 10^{8} M_{\odot}$, followed by testing the FoF finder on this sample by separately applying the 2dFGRS ($b_{J} = 19.45$) and GAMA G23 ($i = 19.2$) magnitude limits. The simulations provided galaxy groups, which were then used as a baseline for calibrating the {\tt FoFpy} algorithm, improving its performance. This process enabled us to evaluate the algorithm's ability to reproduce galaxy groups within the two survey fields, and through these trials, we determined the optimal parameters so that both fields could be used in tandem with a single set of parameters.

By iterating this process numerous times, we discovered that the most effective way to implement the {\tt FoFpy} package was to conduct two separate runs. The initial run designated the ``1$^{\textnormal{st}}$ Pass'', would run on the entire dataset using less constrained limits to extract larger gravitationally bound groups as they could extend further in their dynamical ranges. Using the outputs of the 1$^{\textnormal{st}}$ Pass mode, we filtered out galaxies belonging to groups that contained 10 or more galaxies. The {\tt FoFpy} package would then be re-run on the remaining galaxies, this was designated the ``2$^{\textnormal{nd}}$ Pass''. The 2$^{\textnormal{nd}}$ Pass mode employed stricter constraints to more accurately identify galaxy groups with lower memberships, as this placed better constraints on these less dynamically ranged groups.

\begin{table}[!hbt]
\centering
\caption{Input Parameters of {\tt FoFpy} Python package for the 1$^{st}$ \& 2$^{nd}$ Pass.}
{\tablefont\begin{tabularx}{\columnwidth}{ C C C } 
\hline  \hline
Parameter  &1$^{st}$ Pass    &2$^{nd}$ Pass  \\
\hline
D$_{\rm initial}$ (Mpc)             &0.075    &0.05   \\
D$_{\rm final}$ (Mpc)               &0.275    &0.25   \\
v$_{\rm initial}$ (km s$^{-1}$)     &75       &50     \\
v$_{\rm final}$ (km s$^{-1}$)       &800      &550    \\
D$_{\rm max}$ (Mpc)                 &0.95     &0.75   \\
v$_{\rm max}$ (km s$^{-1}$)         &3400     &1700   \\
$n_{\rm trials}$                    &200      &200    \\
Cutoff                              &0.4      &0.35   \\
\hline  \hline
\end{tabularx}}
\label{FoFpy Parameters}
\end{table}

The input parameters for both the 1$^{\textnormal{st}}$ and 2$^{\textnormal{nd}}$ Passes can be found in Table \ref{FoFpy Parameters}. We then applied the FoF finder to the entire sample, producing a more comprehensive view of the local universe and identifying more accurate galaxy groups compared to using a mass or magnitude-limited sample, which can introduce errors such as the fragmentation of groups. This approach was significantly supported by the 3D visualisation tool {\tt Partiview} \citep{Levy10}, allowing us to visually inspect the groups for quality control and parameter validation by reviewing the links between group members. This analysis also enabled direct comparison with mock datasets, ensuring our linking lengths matched those in the simulated groups.


\subsection{Group Completeness}
\label{subsec:Group Completeness}

The spatial distribution within the designated research area exhibits inhomogeneity arising from variations in survey completeness. Within the 2dFGRS volume chosen for this study, the high completeness of the GAMA G23 volume introduces this inhomogeneity. The GAMA G23 region ($338.1<$RA$<351.9$ and $-35.0<$Dec$<-30.0$) demonstrates notably higher completeness out to $z=0.1$. This volume encompasses a total of 5,844 galaxies, however, upon excluding those exclusively detected by GAMA, the remaining sample contains 2,515 galaxies. Consequently, within the G23 volume of our sample, the completeness, augmented by GAMA, is approximately 2.3 times higher than the remainder of the SGP volume. This discrepancy is particularly pronounced closer to our redshift limit of 0.1, but increasing becomes a noteworthy effect beyond $z>0.06$.

\begin{figure}[!hbt]
\centering
\includegraphics[width=\linewidth]{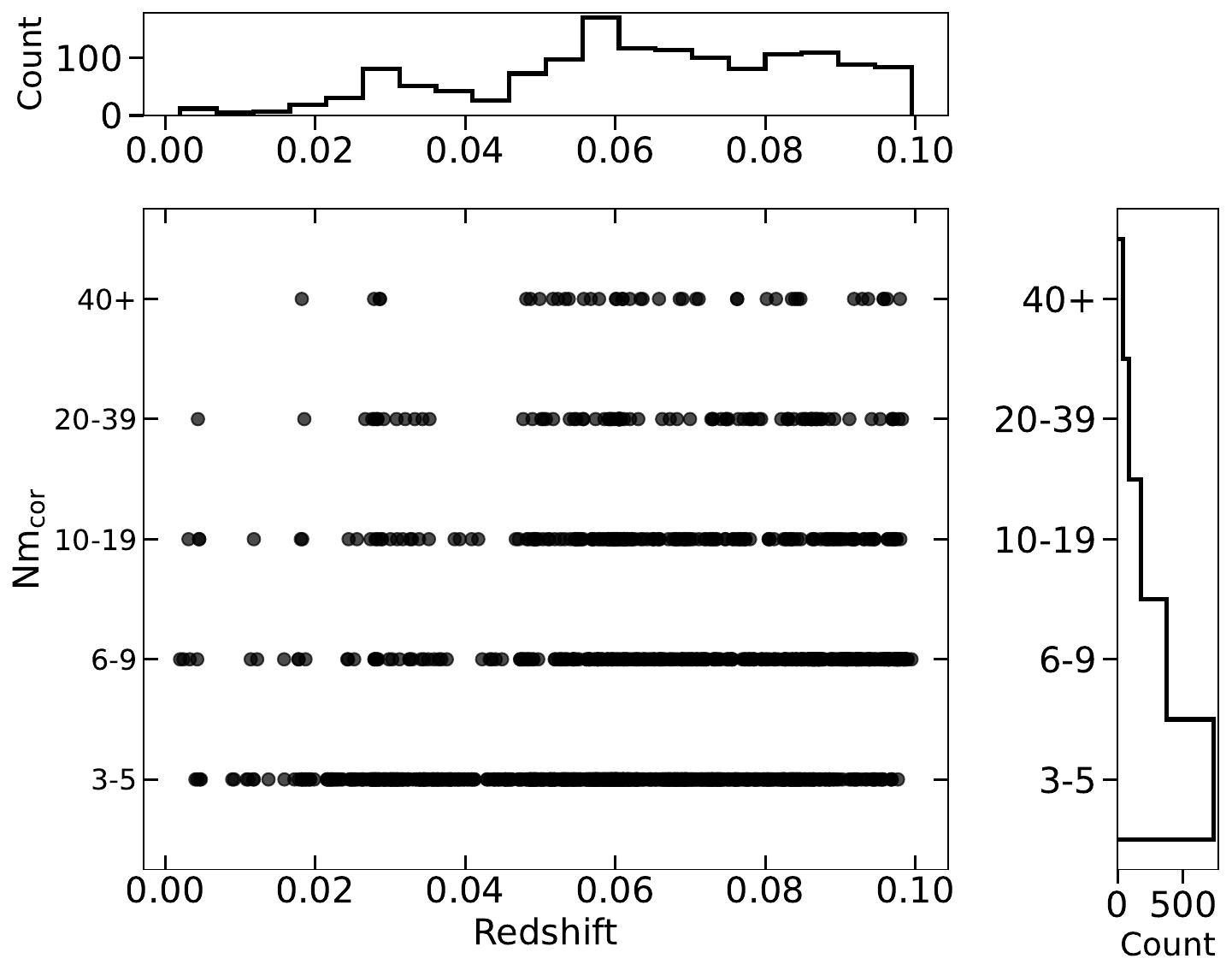}
\caption{Halo occupation distribution (HOD) bins plotted as a function of redshift for the galaxy group sample within the redshift range $z < 0.1$. The HOD bins are separated into five ranges: 3–5, 6–9, 10–19, 20–39, and 40+ corrected members.}
\label{Group_Red}
\end{figure}

As the spectroscopic surveys utilised are magnitude-limited, some galaxies inherently exist beyond these magnitude limits and are not accounted for in these surveys. To address this issue, we again utilise {\tt Millenium} simulations running SAGE to categorise the completeness of our groups across the redshift range. By having the initial simulation with a stellar mass cut of $M_{\star} > 10^{8} M_{\odot}$, ensuring that all galaxies can be adequately resolved from the surrounding dark matter in the SAGE simulation and that the galaxies would make significant physical contributions to the ``local environment''. We then impose a 2dFGRS magnitude limit of $b_{J} = 19.45$ (the main magnitude limit for the SGP region) and contrast it to the GAMA G23 magnitude limit of $i = 19.2$. By comparing group memberships from simulations before and after applying magnitude limits across 25 simulated light cones, we established scaling relationships that increase the number of galaxies within a group as a function of redshift and the observed group membership, providing a more accurate quantification of environment and the underlying dark matter halo occupation. The scaling parameter ($\alpha$) is simply a coefficient to multiply the group membership total to obtain a corrected membership total. For example, if a galaxy group was observed to have 8 members and $\alpha$ was 1.5 due to its redshift position, the ``corrected number of members'' (Nm$_{\rm cor}$) would be 12. Figure \ref{Group_Completeness} shows the relation between $\alpha$ and redshift for different group memberships for each 2dFGRS and GAMA G23. This informs about the expected number of galaxies missing from each group and their relative uncertainties. The equations for the scaling parameters established in Figure \ref{Group_Completeness_Sims} can be found in Appendix \ref{sec:Appendix alpha} in Table \ref{Group Completeness Equations}.

\begin{table}[!hbt]
\centering
\caption{Local environment sample numbers and cross-match statistics. The groups refer to the group membership after the group membership correction has been applied.}
{\tablefont\begin{tabularx}{\columnwidth}{ C C C C }  
\hline  \hline
Local           &No. of Each       &No. of       &No. of\\
Environment     &Environment       &Galaxies     &WISE Matches\\
\hline
Galaxy Groups       &1413    &8980     &8454\\
Groups: 3-5         &733     &2431     &2241\\
Groups: 6-9         &372     &1664     &1579\\
Groups: 10-19       &179     &1526     &1435\\
Groups: 20-39       &88      &1479     &1402\\
Groups: 40+         &41      &1862     &1798\\
Close Pairs         &337     &674      &673\\
Field               &--      &15469    &14055\\
\hline  \hline
\end{tabularx}}
\label{Local Env Numbers}
\end{table}

Insights gleaned from the simulations reveal that low membership groups (3-5 members, 6-9 members, and 10-19 members) necessitate smaller correction factors, albeit with significantly larger uncertainties, particularly at higher redshifts. Conversely, larger membership groups (20-39 members and 40+ members) require more substantial correction factors but exhibit smaller uncertainties, thus affording greater precision in corrections up to higher redshifts. These tendencies are notably accentuated within the 2dFGRS field when compared to the GAMA field. This distinction arises due to the superior completeness of the GAMA survey, which results in comparatively well-constrained correction factors for the ``smaller'' membership groups.

\begin{figure*}[!hbt]
\centering
    \begin{minipage}[t]{\textwidth}
        \centering
        \includegraphics[width=\textwidth]{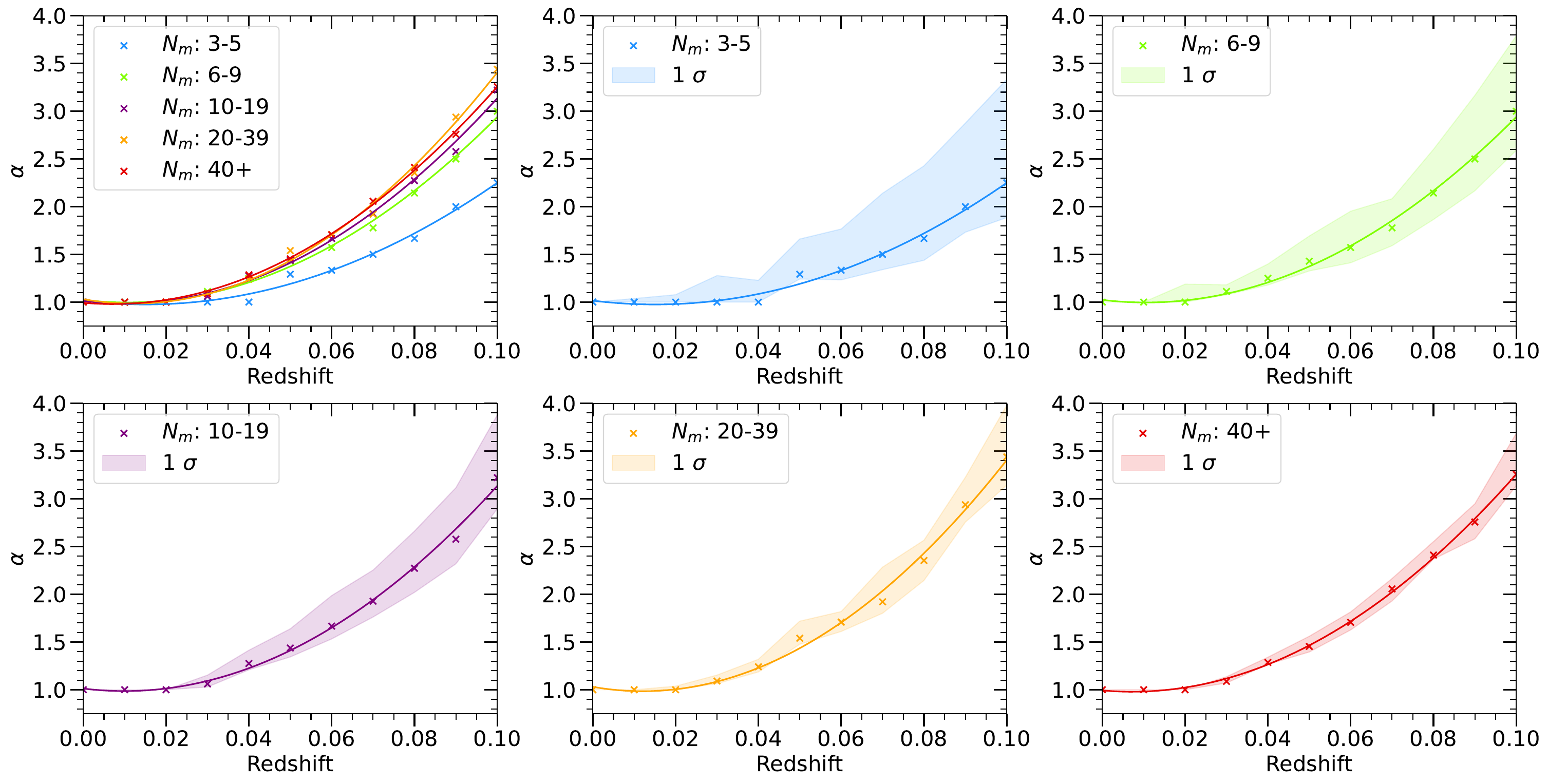}
        \par\vspace{0.5em}
        \parbox{\textwidth}{\centering \small (a) 2dFGRS Group Multiplicity Membership Completeness Scale} \label{2dF_Cor}
    \end{minipage}\vfill
    \begin{minipage}[t]{\textwidth}
        \centering
        \includegraphics[width=\textwidth]{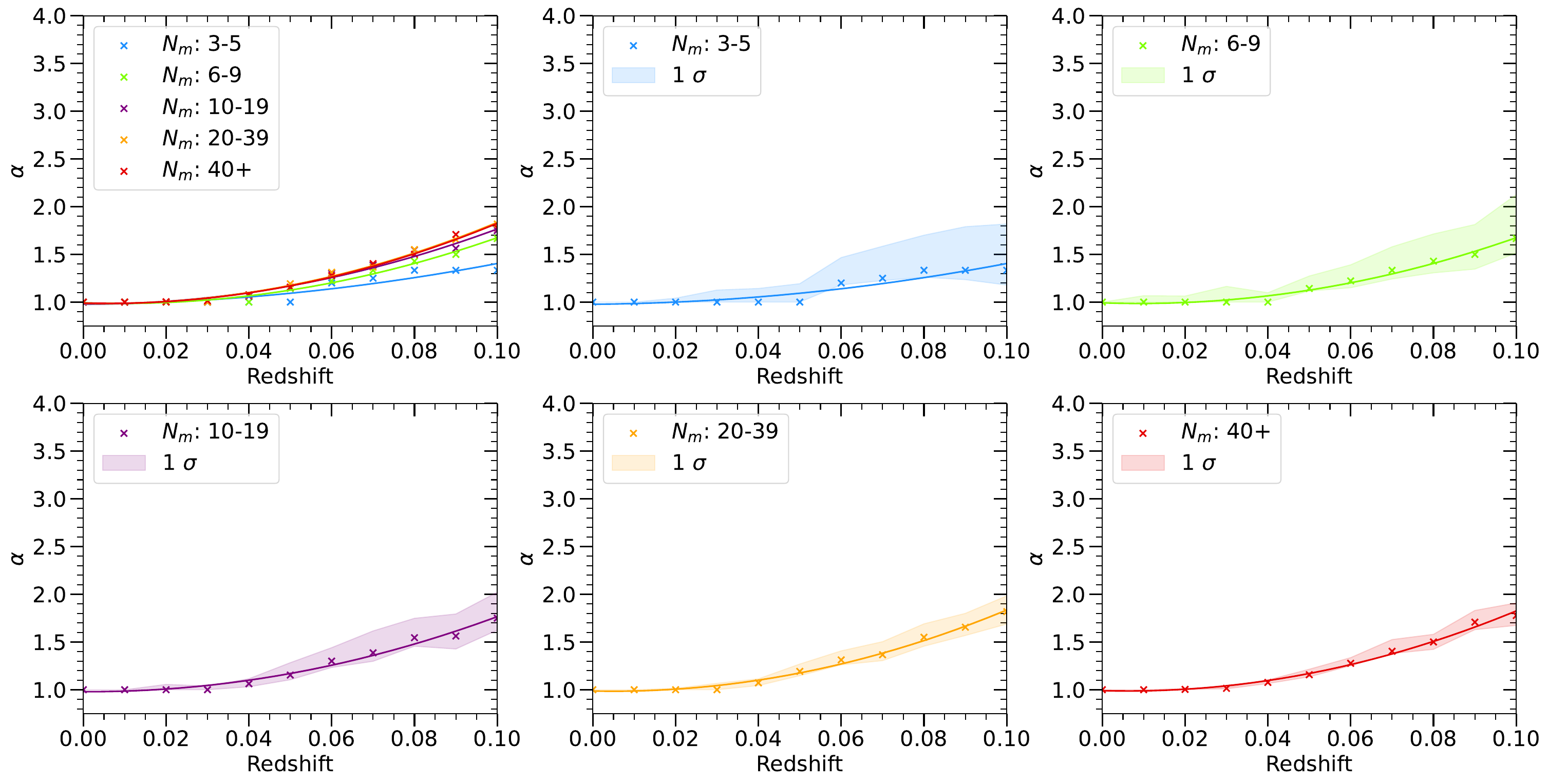}
        \par\vspace{0.5em}
        \parbox{\textwidth}{\centering \small (b) GAMA G23 Group Multiplicity Membership Completeness Scale} \label{GAMA_Cor}
    \end{minipage}
\caption{Comparison of the multiplicity correction factor ($\alpha$) for group membership (as defined in Section \ref{subsec:Group Completeness}) versus redshift for different galaxy group membership bins. Panel (a) illustrates this relation for the 2dFGRS field, where the correction factors are notably larger, particularly at redshifts near our limit. This dataset also exhibits significantly more variability, as evidenced by the shaded 1$\sigma$ regions. In contrast, panel (b) displays the corresponding data for the GAMA G23 field.}
\label{Group_Completeness_Sims}
\end{figure*}

It is evident that as we extend our analysis to larger redshifts, the uncertainties associated with smaller group memberships within the 2dFGRS field become pronounced. For instance, at $z\sim 0.06$, the uncertainties for 3-5 member groups in the 2dFGRS field match those of the GAMA field at $z=0.1$. Similarly, at $z\sim 0.07$, the uncertainties for 6-9 member groups in the 2dFGRS field converge with those of the GAMA field at $z=0.1$, and at $z\sim 0.08$, the uncertainties for 10-19 member groups in the 2dFGRS field align with those of the GAMA field at $z=0.1$. This underscores the importance of survey completeness on the reliability and precision of correction factors, especially for smaller group memberships at larger redshifts.

This analysis enables us to make more accurate assessments and interpretations of the galaxy population within the SGP region and make a statistically more accurate quantification of environment, compensating for the potential incompleteness arising from survey magnitude limits. This, in turn, allows us to scale our galaxy groups to these relations in the various fields, and the effects can be seen in Figure \ref{Group_Completeness}. This is a useful exercise in preparation for more comprehensive spectroscopic surveys such as the upcoming 4-metre Multi-Object Spectrograph Telescope Hemisphere Survey \cite[4HS;][]{ENTaylor23}, which will survey the whole southern sky in the local ($z<0.1$) universe, but with varying completeness in some areas. By understanding how to control and handle varying survey completeness we will be able to maximise the science that can be produced with the spectroscopic information and minimise the error.

The sample is systematically partitioned to categorise galaxies based on their respective small-scale environment. The decision to separate the group sample into five halo occupation distributions (HOD) was made to ensure a relatively even distribution of numbers across the bins whilst representing progressively larger structures. We have explored the utilisation of alternate HOD bins, this did not affect the overall results. Thus our choice of bins balanced the need of maintaining robust statistics, minimising errors while preserving the structural progression of small-scale environments. In Figure \ref{Group_Red}, we demonstrate the HOD distribution as a function of redshift.  Table \ref{Local Env Numbers} provides a breakdown of the total number of galaxies in each small-scale environment and specifies the count of galaxies that have been cross-matched with WISE.

In the context of this analysis, the term ``field'' is operationally defined to encompass galaxies that are not affiliated with either a group or close pair environment. These field galaxies serve as the control within the sample, as they lack association to any small-scale environment.

\subsection{Group Membership - Halo Mass Relation}
\label{subsec:Halo_Mass}

Throughout our analysis, we have adopted group membership as a surrogate for the scale of environmental for the galaxy groups. This choice stems from the considerable uncertainty surrounding halo mass estimations for low membership groups. Typically, observational methods rely on the virial theorem, which establishes a relationship between the group velocity dispersion and virial mass. However, for low membership galaxy groups, this relationship can exhibit significant variability, as further discussed in Appendix \ref{sec:Appendix Halo Mass Var}. Our investigation into various observational techniques for determining halo mass has revealed deviations exceeding 2 dex for low membership groups when compared to our simulated data discussed in Sections \ref{subsec:FoF Algorithm} and \ref{subsec:Group Completeness}. This discrepancy is also acknowledged by \cite{Hess13}, who describe group membership as the HOD and propose that it offers a more stable alternative to halo mass estimation, particularly for low membership groups.

This bias in halo mass estimation can be attributed to the reliance of the virial theorem on gravitationally stable systems, a condition not always met by low membership groups. Additionally, the variance in peculiar velocities further complicates the derivation of a reliable halo mass estimate based solely on velocity dispersion. Given these inherent uncertainties, we have chosen group membership as a more robust and intuitive indicator of the galactic environment rather than relying directly on halo mass. This approach is particularly pertinent to our focus on low-mass groups.

\begin{table}[!hbt]
\caption{Relation between halo mass and group membership HOD. The halo masses are derived from the Millennium-SAGE simulations, with the mean and 1 $\sigma$ values obtained from the histogram distribution in Figure \ref{M200 Hist}.}
\centering
{\tablefont\begin{tabularx}{\columnwidth}{ CC }
\hline \hline 
Number of Members  & $\log M_{200} \: (M_{\odot})$  \\ 
\hline
3 - 5             & 12.31$\pm$0.49   \\
6 - 9             & 12.87$\pm$0.36   \\
10 - 19           & 13.23$\pm$0.31   \\
20 - 39           & 13.59$\pm$0.27   \\
40 +              & 13.96$\pm$0.32   \\
\hline \hline
\end{tabularx}}
\label{Halo Mass - Members}
\end{table}

It is worth noting that there exists a correlation between group membership and halo mass, as evidenced in Table \ref{Halo Mass - Members}, where we present the mean halo mass given by $M_{200}$ ($M_{\odot}$) derived from the SAGE simulations for each galaxy group membership bin employed in our analysis. This table utilises all the light cones outlined in Section \ref{subsec:Group Completeness} and provides the mean values of halo mass for our chosen HODs. Additionally, we provide the $1 \sigma$ variance of these binned halo masses. The distribution of these halo masses can be found in Figure \ref{Delta M200}. It is worth noting that the local group falls between the expected halo masses of the 3-5 and 6-9 member groups, but closer to the 3-5 groups with a halo mass of $\sim \log 12.5 \: (M_{\odot})$ \cite{Sawala23}.


\subsection{Pair Selection}
\label{subsec:Pair Selection}

In addition to galaxy groups, we specifically examine close galaxy pairs to investigate the behaviour of interacting galaxies. Close pairs were chosen as the best method to study closely interacting systems, acknowledging the inherent challenges such as projection effects, the inclusion of pairs within larger groups, and potential completeness issues near our redshift limit. We do not aim to define a state-of-the-art pair sample; rather, our goal is to gather a selection of closely interacting galaxies to explore how extreme interactions and probable future mergers influence star formation relative to galaxies in the field and within groups, and whether these potential pre-mergers exhibit enhanced environmental effects seen in group environments.

Close pairs were defined as galaxies with a projected spatial separation of $r_{\rm sep} < 50 : \text{h}^{-1} \text{kpc}$ and a radial velocity difference of $v_{\rm sep} < 500 : \text{km s}^{-1}$, ensuring no additional galaxies met this criterion for each pair. This follows the works of \cite{Robotham2014}, \cite{Bok2020}, and \cite{Contreras-Santos2022}, who identified it as an optimal pair selection method. Despite the separation criteria, the linking lengths for galaxy groups (see Section \ref{subsec:FoF Algorithm}) can extend beyond these limits, resulting in some close pairs also being members of larger groups. Additionally, close pairs near the redshift limit may include neighbouring galaxies below the detection threshold, leading to potential misclassifications.

Identifying galaxy pairs presents challenges, particularly due to variations in peculiar velocities within group environments. According to \cite{Contreras-Santos2022}, our selection criteria accurately identify galaxy pairs in approximately 50-60\% of cases across the angular separation range. Nonetheless, this method remains one of the most reliable for producing pair samples in large surveys. We do not exclude pairs near the redshift limit or those within groups, as our focus is on studying the effects of close interactions rather than ensuring a pure pair sample. This approach also enables a more direct comparison with previous studies that do not separate or exclude pairs within groups \citep[e.g.][]{Lambas03, Ellison10, Woods10, Scudder12, Hopkins2013, Robotham2014, Davies15, Bok2020, Sun2020, Steffen21, Shah2022, Li23}, especially given the high uncertainty in correctly associating galaxy pairs. As shown in Table \ref{Local Env Numbers}, we have 674 galaxies within close pairs and 673 have been cross-matched with WISE. Of the 673 cross-matched pairs, 449 pairs are in groups, and 224 are ``isolated'' and not part of any group.

All candidate pairs underwent visual confirmation using WISE and DESI Legacy Imaging Surveys \citep{Dey19}, providing better constraints on our selection. However, WISE data posed limitations for pairs with angular separations smaller than 6", roughly the beam width of W1, making it difficult to resolve secondary galaxies. Such pairs were excluded from further analysis. For pairs with separations between 6" and 10", resolution depended on the relative sizes of the galaxies, while separations beyond 10" were consistently well-resolved in WISE. At $z = 0.05$, a separation of 6" corresponds to $8.8 \: \text{h}^{-1}$ kpc, and 10" corresponds to $14.7 \: \text{h}^{-1}$ kpc.

\section{Analysis}
\label{sec:Analysis}

In this analysis, we examine how the quenched fraction and SF of galaxies vary across small-scale environments (Sections \ref{subsec:SFQ_Local} and \ref{subsec:SF_Local}), specifically focusing on the field, close pairs, and galaxy groups. By analysing these small-scale environments, we aim to better understand how galaxy interactions and group dynamics influence SF across different environments. Additionally, we explore how the quenched fraction and SF within groups vary as a function of HOD (Sections \ref{subsec:SFQ_Group} and \ref{subsec:SF_Group}) to assess how the growth of small-scale environments impacts these properties. Finally, we investigate how SF properties change based on the mass ratios of interacting close pairs, and test for induced SF as a function of projected separation (Section \ref{subsec:Close Pair Effects on Star Formation}).

\subsection{Star Formation Quenching in Local Galaxy Environments}
\label{subsec:SFQ_Local}


In the subsequent analysis, we will employ the quenching line of $\log sSFR = -11.0$, as defined in Section \ref{subsubsec:Star-Forming Main Sequence}, to identify quenched galaxies and quantify this as a quenched fraction. The quenched fraction is calculated as the ratio of quenched galaxies to the total number of galaxies within a specified sample and stellar mass range. Through analysis across various stellar mass bins, we aim to investigate the environmental dependence of SF quenching through the evolving quenched fraction across different local environments.

\begin{figure}[!hbt]
\centering
\includegraphics[width=\linewidth]{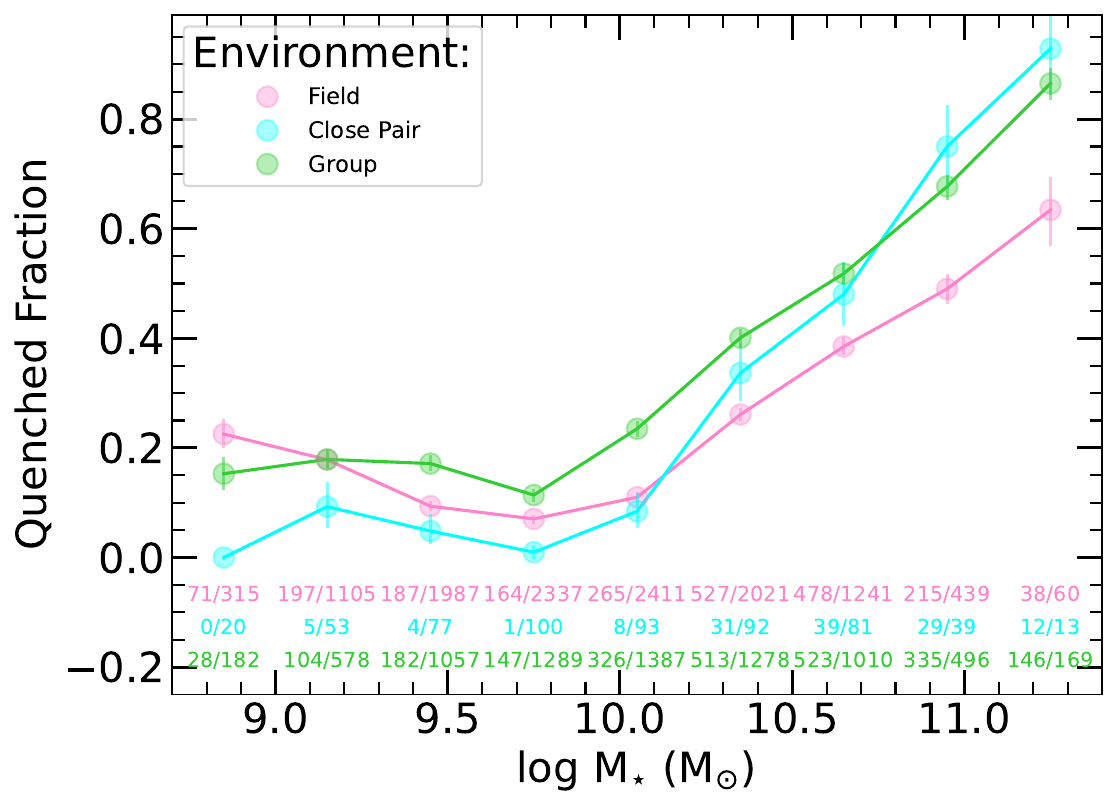}
\caption{The fraction of quenched galaxies per given stellar mass bin across various local galaxy environments. The quenched fraction represents the ratio of quenched galaxies to the total number of galaxies per mass bin. Each mass bin has a width of 0.3 dex, with errors calculated via bootstrap resampling within each bin. The total number of quenched galaxies and the total number of galaxies for each mass bin are provided below the distributions. The quenched fraction's evolution is observed across different environments and stellar masses.}
\label{Local_QF}
\end{figure}

\begin{figure*}[!hbt]
\centering
\includegraphics[width=\linewidth]{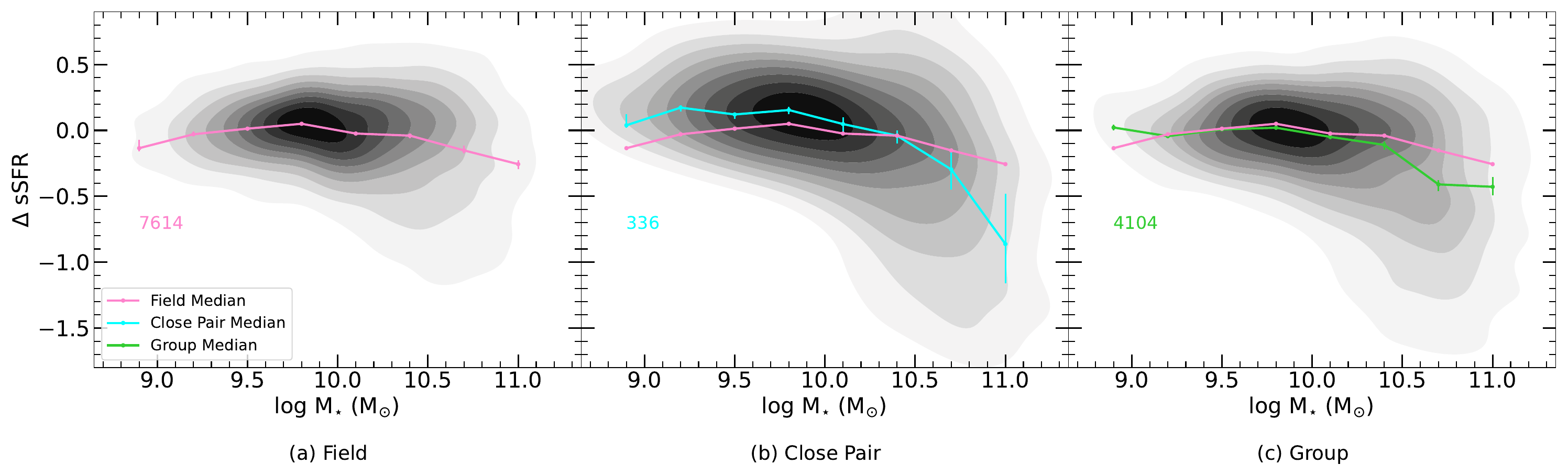}
\caption{The relationship between $\Delta sSFR$ and stellar mass across various local galaxy environments is depicted. Contours in each panel illustrate the distribution of all non-SFR UL and SFR S/N $>$ 2 galaxies within the mass-complete sample, delineating the SF population. The median values of each distribution are depicted alongside the field galaxy median distribution in pink within each panel to highlight the relative environmental effects. A $\Delta sSFR > 0$ is more efficiently SF than the SFMS fit in Figure \ref{SFMS} and $\Delta sSFR < 0$ is less efficiently SF. Median values are binned in stellar mass bins of 0.3 dex, with errors computed via bootstrap resampling within each bin. The total number of sources used in the median distributions, i.e. non SFR-UL and SFR S/N $>$ 2 sources, is indicated in the lower left with the same colouring as the median distribution. The close pair distribution shows an increase in SF activity at most masses, with a relative decrease at high masses. Group galaxies exhibit relative similarities to the field, except at low and high masses where they show similarities to close pairs.}
\label{Delta_Local}
\end{figure*}

Given the robustness of our constructed sSFR-SM relations, this section of the analysis will incorporate our mass-complete sample (see Section \ref{subsubsec:Mass Comp}, including IR warm sources and low-S/N. SFR UL values, provided they adhere to the criteria outlined in \cite{Cluver2020} are included here for completeness. SFR ULs typically arise from the W3 flux being below the detectable threshold. These sources are checked to see if the W3 flux (given the distance to the source) is above or below this threshold. UL values above this threshold are included within the analysis of the quenched fractions and correlate to a quiescent galaxy, those below the threshold are treated conservatively and are included as a star-forming galaxy within the binning, so as not to boost the quenched fraction artificially. For more details on how UL were treated in the analysis see Appendix \ref{sec:Appendix UL}.

In Figure \ref{Local_QF}, we scrutinise the quenched fraction across different local environments, encompassing ``field'' galaxies, close pairs, and all galaxies within groups, thereby spanning a continuum of low-density to high-density small-scale environments. Figure \ref{Local_QF}, along with subsequent figures in this analysis, have their associated uncertainties derived using bootstrap resampling. This method involves repeatedly sampling the data with replacement to estimate the variability and confidence intervals for the observed metrics.

Our findings show a changing quenched fraction across various stellar mass bins. Specifically, the stellar mass bin  $M_{\star} = 10^{9.75} \: M_{\odot}$ exhibits the lowest quenched fraction for all local environments, indicative of the highest efficiency of SF within galaxies across all mass ranges. Conversely, for mass bins with $M_{\star} < 10^{9.75} \: M_{\odot}$, we observe a gradual increase in the quenched fraction as we descend to lower stellar masses. This trend suggests that SF in these low-massed galaxies is less efficient. This is particularly evident in smaller dwarf galaxies which exhibit lower efficiency in converting available gas into stars \citep{Moster13,Hunt20}.

In the larger stellar mass range ($M_{\star} > 10^{9.75} \: M_{\odot}$), we observe a consistent rise in the quenched fraction with increasing stellar mass. This observation is to be expected, as galaxies evolve through the combination of in-situ SF and merger processes, evolving into sizable, high-mass galaxies, which result in SF inefficiencies or global SF quenching due to mass-related quenching mechanisms \citep{Gabor2010,Peng10,Bluck20,ETaylor23}.

Contrasting the close pairs to the field population, at low stellar masses ($M_{\star} < 10^{9.75} \: M_{\odot}$), the close pairs demonstrate a lower quenched fraction to their field counterparts, which suggests that low-mass close pairs exhibit more efficient SF. This result has been observed in prior observational studies, such as those by \cite{Hopkins2013} and \cite{Sun2020}, where gravitational and tidal interactions between paired galaxies facilitate the funnelling of gas into central regions, effectively constraining gas outflows, fostering conditions conducive to central SF. These effects in close pairs synthesise in an overall increase to the galaxy's SFR \citep{Ellison10}. However, our analysis encounters limitations in the lowest mass bin due to diminished statistical significance, underscoring the need for a larger sample size to probe this mass range's aggregate behaviour effectively.

In contrast, at higher stellar masses ($M_{\star} > 10^{9.75} \: M_{\odot}$), close pairs exhibit a higher degree of suppression compared to their field counterparts, suggesting a potential relationship between the close pair environment and SF quenching in higher-mass galaxies. This will be examined more in-depth within Section \ref{subsec:Close Pair Effects on Star Formation}. Additionally, Appendix \ref{sec:Appendix Pairs} explores the differences between close pairs within and outside groups, where we observe variation between the two at $M_{\star} > 10^{10} \: M_{\odot}$. However, due to the limited sample size and uncertainty in pair classifications, these differences are not statistically significant. For this reason, we maintain a unified close pair sample in the main analysis to ensure consistency with prior research \citep[e.g.][]{Lambas03, Ellison10, Woods10, Scudder12, Hopkins2013, Robotham2014, Davies15, Bok2020, Sun2020, Steffen21, Shah2022, Li23}. Future studies, with more complete data from upcoming surveys such as 4HS \citep{ENTaylor23}, could further investigate the distinct behaviours of close pairs in and out of groups, offering a more controlled definition of close pairs.

The quenched fraction of the grouped galaxies in the low stellar mass range ($M_{\star} < 10^{9.75} \: M_{\odot}$) is more closely aligned with the field, albeit with a slight increase in the quenched fraction. These findings indicate that in the low-mass regime, group galaxies closely resemble their field counterparts, with a minor impact on their quenched fraction. This observed relation diverges at the higher mass end ($M_{\star} > 10^{9.75} \: M_{\odot}$), where a significant offset between the quenched fraction of the field and grouped galaxies exists. This observation is in agreement with previous works \citep[e.g.][]{Davies19,Cluver2020,Contini20,Delgado22}, which also show larger SF suppression in galaxy groups than in the field, which also is in agreement with the morphology density relation \citep[e.g.][]{Dressler97,Capak07}.

\subsection{Star Formation in Local Galaxy Environments}
\label{subsec:SF_Local}


We next explore changes within only the SF population of galaxies as a function of local environment.  This is done in the form of $\Delta sSFR$, the dex difference between the sSFR measurement and the sSFR from the SFMS polynomial fit of Equation \ref{SFMS_Equation} ($\Delta sSFR = \log sSFR - \log sSFR_{fit}$). Thus a galaxy with $\Delta sSFR = 0$ lies directly on the SFMS fit, a galaxy with $\Delta sSFR > 0$ is more efficiently star-forming than the SFMS fit and $\Delta sSFR < 0$ is less efficiently star-forming. Hence, a change in $\Delta sSFR$ indicates an increase or decrease in SF efficiency within the star-forming population.

In Sections \ref{subsec:SF_Local} and \ref{subsec:SF_Group} the analysis of the $\Delta sSFR$ is primarily qualitative, serving to highlight any variations between different small-scale environments while controlling for stellar mass. This approach facilitates an initial exploration of environmental effects to SF. In Section \ref{subsec:Discussion SFD}, we further refine this analysis and discuss the impact of environment by introducing a new metric (\textit{environmental star formation deficiency}) that quantifies the fractional excess of $\Delta sSFR$ across these small-scale environments, relative to the field, and subject it to a quantitative statistical evaluation.

To analyse only the SF population of galaxies, we do not include SFR ULs; as previously outlined in Section \ref{subsec:SFQ_Local} and Appendix \ref{sec:Appendix UL} these indicate a quiescent galaxy (given they are above the detectable W3 threshold). We do not include measurements with a SFR S/N $<$ 2 within this analysis, as their position along the $\Delta sSFR$ plane is highly uncertain. This selection allows the inclusion of galaxies with $\log sSFR < -11.0$, as this is quite a conservative quenching separator, especially for high-massed galaxies. Not all galaxies with $\log \text{sSFR} < -11.0$ at higher stellar masses are fully quiescent; rather, they are significantly less efficient compared to their lower-mass counterparts. The $\Delta sSFR$ of the SF populations for the different local galaxy environments are contoured in Figure \ref{Delta_Local}. In each panel of Figure \ref{Delta_Local}, the running median of the field $\Delta sSFR$ distribution is plotted by the pink distribution and contrasted with the close pairs in cyan in the middle panel, and groups in green on the right panel.


In contrasting the close galaxy pairs to the field distribution, the close pairs exhibit heightened SF efficiency in the star-forming population at lower stellar masses (seen by an increase in $\Delta sSFR$), suggesting enhanced interactions or gas accretion processes. As stellar mass increases within the close pairs, the SF efficiency observed at lower masses decreases relative to the field, becoming less efficient than the field population at $M_{\star} \sim 10^{10.4} \: M_{\odot}$. In the largest mass bin, the close pairs exhibit significantly larger SF inefficiencies compared to the field (seen by a relative decrease in $\Delta sSFR$). Among the close pair sample, we observe that low-mass galaxies exhibit few quenched members (Figure \ref{Local_QF}), accompanied by an enhancement in $\Delta sSFR$, suggesting an increase in star formation activity. Conversely, at higher stellar masses, we find both elevated quenched fractions (Figure \ref{Local_QF}) and suppressed $\Delta sSFR$, indicating these galaxies are undergoing quenching processes. This contrast suggests that while low-mass galaxies in pairs may experience star formation enhancement, higher-mass pairs are more likely progressing towards quenching.

The combined galaxy group distribution closely resembles the field environment, particularly throughout the low mass regime ($M_{\star} < 10^{9.75} \: M_{\odot}$), except for the lowest mass bin, where we see an increase in $\Delta sSFR$ compared to the field. The similarities between the field and galaxy group relations deviate at $M_{\star} > 10^{10.1} \: M_{\odot}$, marked by a gradual downturn in $\Delta sSFR$ for the groups. The offset in $\Delta sSFR$ between the two populations increases with increasing stellar mass. The offsets observed at stellar masses such as $M_{\star} \sim 10^{10.4} \: M_{\odot}$ are unlikely to be attributable to secular processes at this mass range. Instead, they suggest the presence of environmental influences, potentially arising from mechanisms such as galaxy harassment, strangulation, tidal stripping, or ram pressure stripping \citep[e.g.][]{Moore96,Taranu14,Bahe19}.

\subsection{Star Formation Quenching in Group Environments}
\label{subsec:SFQ_Group}

In the following analysis, we will employ the methodologies for the quenched fraction utilised in Section \ref{subsec:SFQ_Local} applying this methodology to the group galaxies. We separate the group sample by placing galaxies into their discretised membership bins ranging from 3-5, 6-9, 10-19, 20-39, and 40+ corrected members as outlined in Section \ref{subsec:Group Completeness}. Our objective is to investigate whether the increase in group membership/small-scale environment impacts SF processes within galaxies.

\begin{figure}[!hbt]
\centering
    \includegraphics[width=\linewidth]{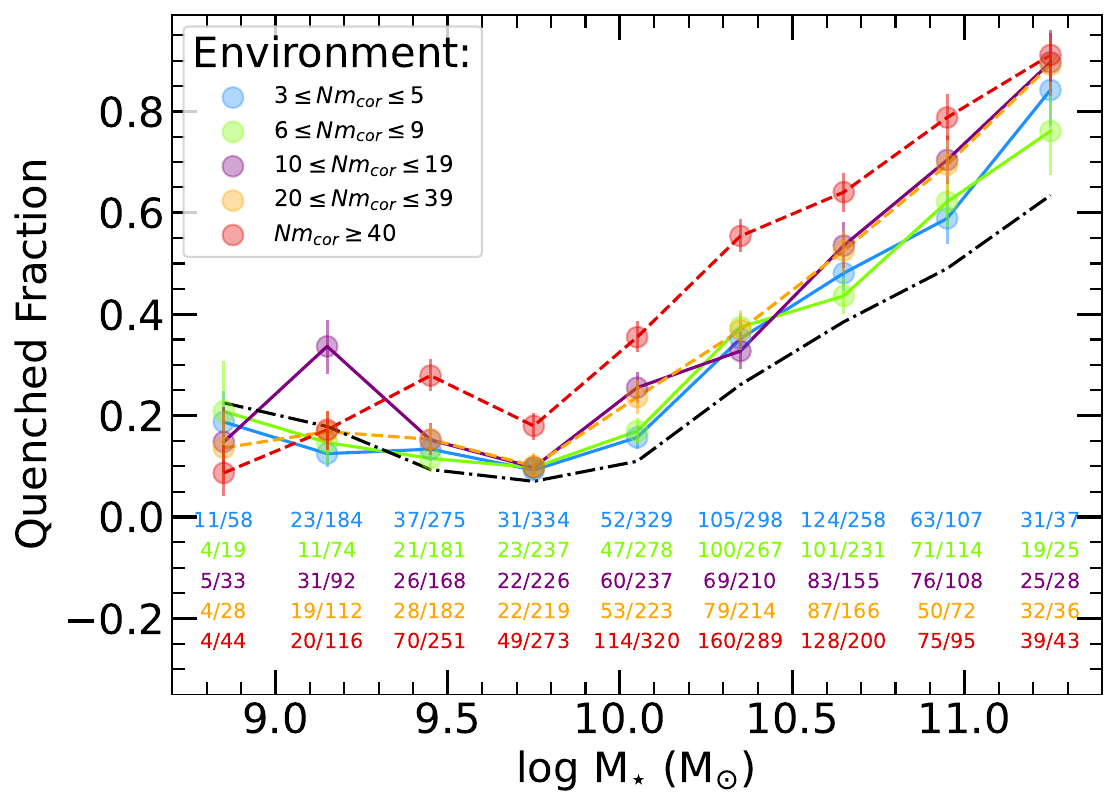}
\caption{The fraction of quenched galaxies within galaxy groups per given stellar mass bin across varying levels of galaxy group membership. The quenched fraction represents the ratio of quenched galaxies to the total number of galaxies per mass bin. The field population quenched fraction is depicted with a dash-dot black line to highlight the relative effects of environment. Each mass bin has a width of 0.3 dex, with errors computed via bootstrap resampling within each bin. The total number of quenched galaxies and the total number of galaxies for each mass bin are provided below the distributions. A noticeable evolution of the quenched fraction is evident across different group environments and stellar masses.}
\label{Group_QF}
\end{figure}

Figure \ref{Group_QF} presents the quenched fraction of the different membership groups as a function of stellar mass to assess the impact of group environments on SF quenching. To contextualise these results, they are juxtaposed with the distribution of the control field sample, represented by the black dot-dashed line initially depicted as the pink line in Figure \ref{Local_QF}. This enables a relative assessment of the impact of growing small-scale group environment on SF quenching processes.

In smaller-membership groups (3-5, 6-9, and 10-19 corrected members), we observe that in the low-mass range ($M_{\star} < 10^{9.75} \: M_{\odot}$), the group environment generally mirrors the trends observed in field galaxies. The 10-19 membership groups in this low mass range demonstrate a slight increase in the quenched fraction compared to the other ``small groups'', but it does not significantly deviate from the field sample until the second lowest mass bin, where a significant increase is observed. Following the increase, it then drops below the field sample within the lowest massed bin. The number statistics for the two lowest and largest massed bins for the various galaxy group sizes in Figure \ref{Group_QF} is substantially less than the rest of the population and thus, the results should be interpreted cautiously due to their significantly lower statistical power, likely necessitating a larger dataset to discern the true underlying trends. 

The observed results are consistent with earlier research conducted by \cite{Kauffmann04} and \cite{Peng10}, which revealed that low-mass galaxies inhabiting environments with low densities or smaller-sized groups tend to display SF levels akin to those of galaxies not associated to groups or residing in even less dense environments.

Conversely, we observe a notable deviation between the group environments and the field distribution when examining the low-mass ($M_{\star} \leq 10^{9.75} M_{\odot}$) range within larger groups (20-39 and 40+ corrected members). Here we see a rise in the quenched fraction, except for the lowest mass bin, where larger groups display a lower fraction compared to the field. As previously discussed, the lowest-massed bin is subject to relatively low statistical significance. However, this result necessitates further investigation to ascertain its significance. The heightened quenched fraction seen elsewhere within the low mass range is particularly evident in the 40+ corrected membership groups and suggests a potential influence of ram pressure stripping on low-mass galaxies as they encounter massive, hot dark matter halos. This mechanism has been previously documented in studies observationally and through simulations \citep[e.g.][]{Kapferer08, Ebeling14, Taranu14, Merluzzi16}, where interactions between galaxies and the ICM result in the removal of neutral gas from galaxies. Consequently, this rapid gas loss leads to inefficient SF or the quenching of SF, especially within the discs of these galaxies. However, if these galaxies are gas-rich, it can result in centralised regions of SF and a net increase in their overall SFR, which can potentially explain our findings in the lowest mass bins.

\begin{figure*}[!hbt]
\centering
\includegraphics[width=\linewidth]{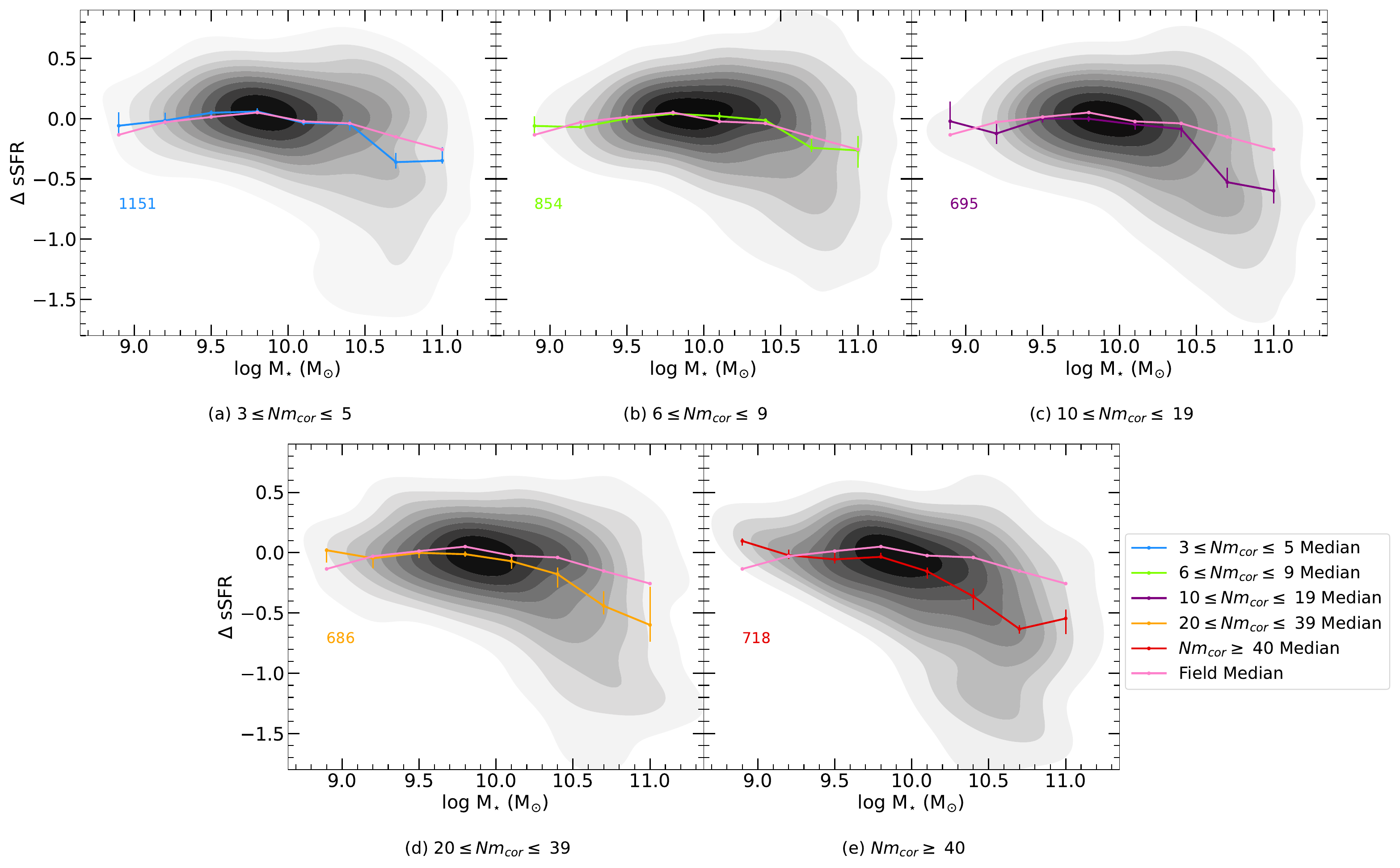}
\caption{The relationship between $\Delta sSFR$ and stellar mass across various galaxy group environments is depicted. Contours in each panel illustrate the distribution of all non-SFR UL and SFR S/N $>$ 2 galaxies within the mass-complete sample, delineating the SF population. Median values of each distribution are depicted alongside the field galaxy median distribution in pink within each panel to highlight the relative environmental effects. A $\Delta sSFR > 0$ is more efficiently SF than the SFMS fit in Figure \ref{SFMS} and $\Delta sSFR < 0$ is less efficiently SF. Median values are binned in stellar mass bins of 0.3 dex, with errors computed via bootstrap resampling within each bin. The total number of sources used in the median distributions, i.e. non SFR-UL and SFR S/N $>$ 2 sources, is indicated in the lower left with the same colouring as the median distribution. The group galaxies show a decrease in their SF at large stellar masses compared to the field, which increases with group membership.}
\label{Delta_Group}
\end{figure*}

In the high mass range ($M_{\star} > 10^{9.75} \: M_{\odot}$), there is a noticeable deviation of the quenched fraction between the group environments and the field distribution across all group sizes. This offset becomes more pronounced as group size increases. Notably, as galaxy groups increase in their membership, larger quenched fractions are observed, indicating heightened effectiveness of environmental SF quenching. This result may be influenced by the varying stellar mass function across different galaxy group environments \citep{Mehmet15, Etherington17, Papovich18}, with larger groups more likely to host massive, ``red and dead'' galaxies. However, this does not seem to drive the observed results. As noted by \cite{Calvi13}, the stellar mass function may not change significantly across field and group environments, which is supported by our finding that the relative number of galaxies in the highest stellar mass bins remains similar across different HODs. In this mass range, mass quenching effects dominate, as shown by the field sample, but are further amplified by environmental processes that become increasingly pronounced with larger group sizes. The impact of the group environments is consistent with previous studies such as \cite{Davies19} and \cite{Cluver2020}, which found an increase in the passive fraction as a function of halo mass when controlling for stellar mass, mirroring our observations due to the relation between halo mass and group size discussed in Section \ref{subsec:Halo_Mass}. Additionally, our results align with those of \cite{Delgado22}, demonstrating that while mass quenching effects dominate in the larger stellar mass range, environmental effects play a significant role in amplifying the suppression of SF.

\subsection{Star Formation in Group Environments}
\label{subsec:SF_Group}

We apply the methodologies outlined in Section \ref{subsec:SF_Local} to our galaxy group sample to investigate the environmental impact on the SF population of galaxies, using the relative variance in $\Delta sSFR$ as the probe. In Figure \ref{Delta_Group}, the running median distribution for each binned group is compared to the field's median distribution, depicted by the pink line and originally established in Figure \ref{Delta_Local}. As described in Section \ref{subsec:SF_Local}, SFR ULs and SFR low S/N measurements are excluded from the contouring and running medians to represent the SF population accurately.


Our findings from Figure \ref{Delta_Group} highlight a discernible evolutionary shift in SF efficiency as galaxy groups increase in size. We observe that the offset in $\Delta sSFR$ between the field and group samples grows with increasing group size, particularly at higher stellar masses.

In 3-5 and 6-9 membership groups, the median SF population typically mirrors the field distribution, except for a slight decrease in $\Delta sSFR$ beyond $10^{10.4} M_{\odot}$. As we increase to 10-19 member groups and larger, there is an increasing offset between the galaxy groups distributions and that of the field. The relative decrease in $\Delta sSFR$ compared to the field in these galaxy groups becomes more pronounced in larger galaxy groups as their membership increases. The initial deviation between the field and the galaxy groups shifts to lower stellar masses as group membership increases. This initial offset begins at $10^{10.4} M_{\odot}$ for 10-19 membership groups and evolves to $10^{9.8} M_{\odot}$ for 40+ membership groups.

In low-membership groups, the star formation SF inefficiencies among higher-mass galaxies are primarily driven by secular processes. As we examine slightly larger groups, while secular quenching mechanisms continue to play a dominant role in suppressing SF, the environmental conditions within these groups further enhance the suppression of SF, leading to the pre-processing of SF within the groups. At the lowest stellar mass bin for all groups, we see an increase in $\Delta sSFR$ relative to the field. This trend may potentially indicate similar interactions experienced by the low-mass close pairs, as further analysed in Section \ref{subsec:Close Pair Effects on Star Formation}.

In summary, our analysis reveals that as galaxy groups increase in membership/HOD, the deviation in SF efficiency from the field population becomes more pronounced, particularly at higher stellar masses, suggesting a complex interplay between stellar mass and environmental effects present in group environments on SF dynamics that we will further discussed in Section \ref{sec:Discussion}.

\subsection{Star Formation in Close Pairs}
\label{subsec:Close Pair Effects on Star Formation}

In this section of our analysis, we drop the use of the redshift-dependent mass complete cut to the sample. This extension is prompted by several considerations, with a key factor being the limitations posed by the low number statistics of our close pairs, which hinders an in-depth analysis while we control for stellar mass. By incorporating all close pairs, we augment our sample by an additional 20\%. Another rationale for this stems from the treatment of our group samples, which involved applying completeness corrections. This step was omitted for the pairs due to the significant variability in results, especially near our redshift limit. The identified close pairs at higher redshifts likely correspond to small galaxy groups. Our aim is not to constrain a highly complete sample of close pairs but rather to investigate the diverse effects resulting from galaxy-galaxy interactions and potential pre-mergers.

\begin{figure}[!hbt]
\centering
\includegraphics[width=\linewidth]{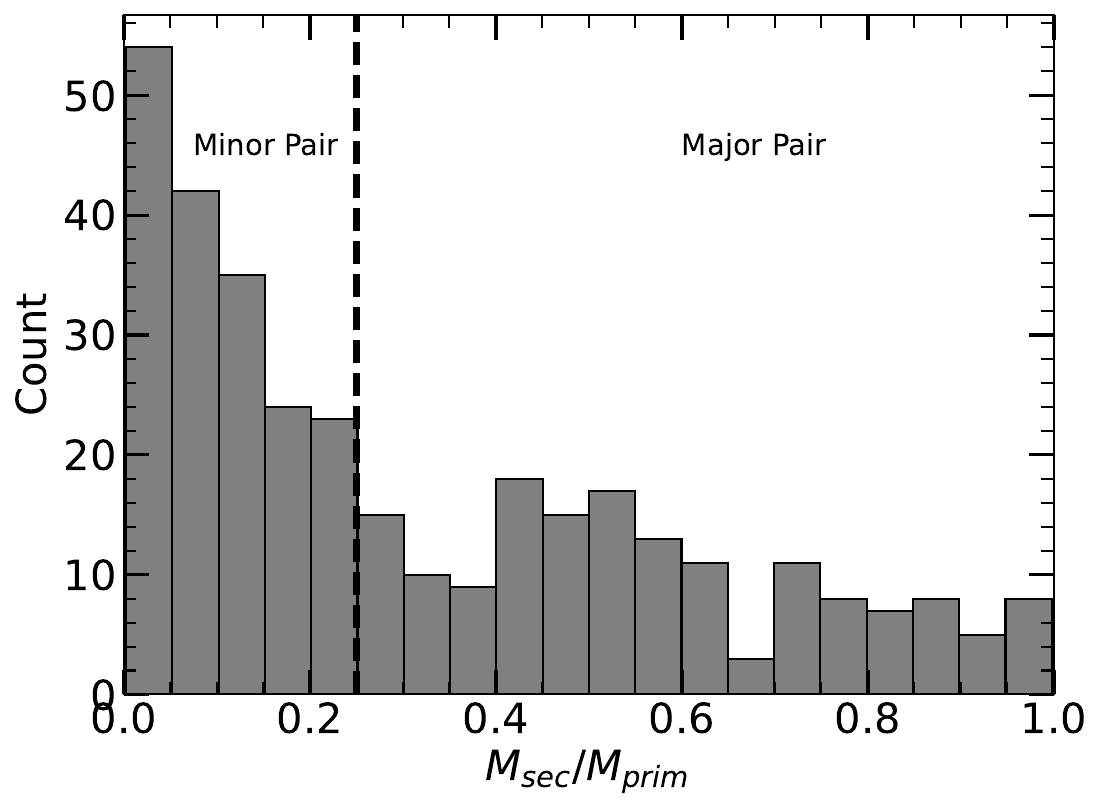}
\caption{Stellar mass ratio of the secondary and primary galaxy of the close pairs. The dashed black line separates those we associate as minor pairs (left of the dashed line) and that of major pairs (right of the dashed line).}
\label{mratio}
\end{figure}

\begin{figure*}[!hbt]
\centering
\includegraphics[width=\linewidth]{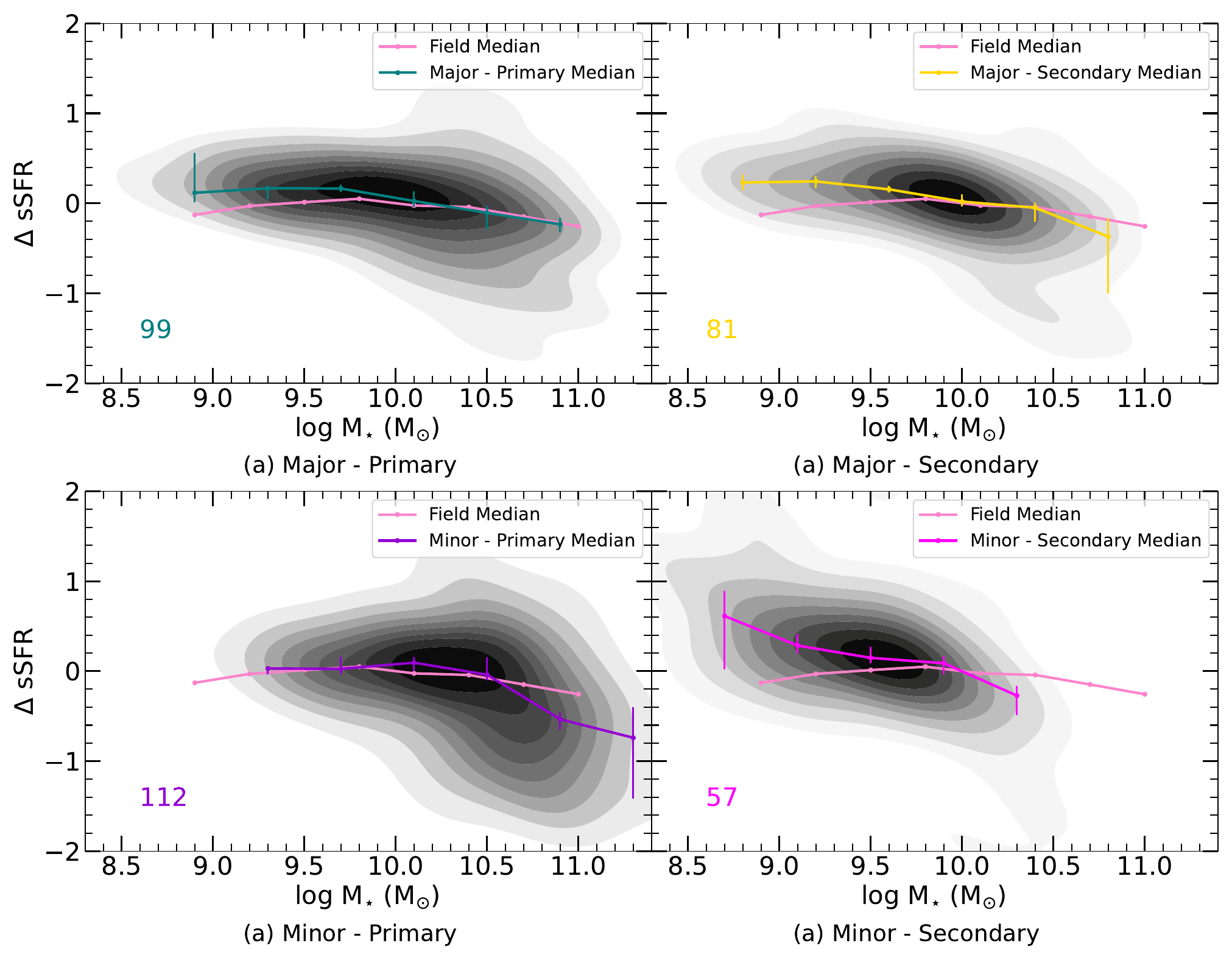}
\caption{The relationship between $\Delta sSFR$ and stellar mass across various close pair environments is depicted. Contours in each panel illustrate the distribution of all non-SFR UL and SFR S/N $>$ 2 galaxies within the mass-complete sample, delineating the SF population. The median values of each distribution are depicted alongside the field galaxy median distribution in pink within each panel to highlight the relative environmental effects. A $\Delta sSFR > 0$ is more efficiently SF than the SFMS fit in Figure \ref{SFMS} and $\Delta sSFR < 0$ is less efficiently SF. Median values are binned in stellar mass bins of 0.4 dex, with errors computed via bootstrap resampling within each bin. The total number of sources used in the median distributions, i.e. non SFR-UL and SFR S/N $>$ 2 sources, is indicated in the lower left with the same colouring as the median distribution. There is a significantly varying SF trend between the major/minor and primary/secondary pairs, ranging from SF enhancements to SF deficiencies across the stellar mass range.}
\label{Delta_Pairs}
\end{figure*}

\begin{figure}[!hbt]
\centering
\includegraphics[width=\linewidth]{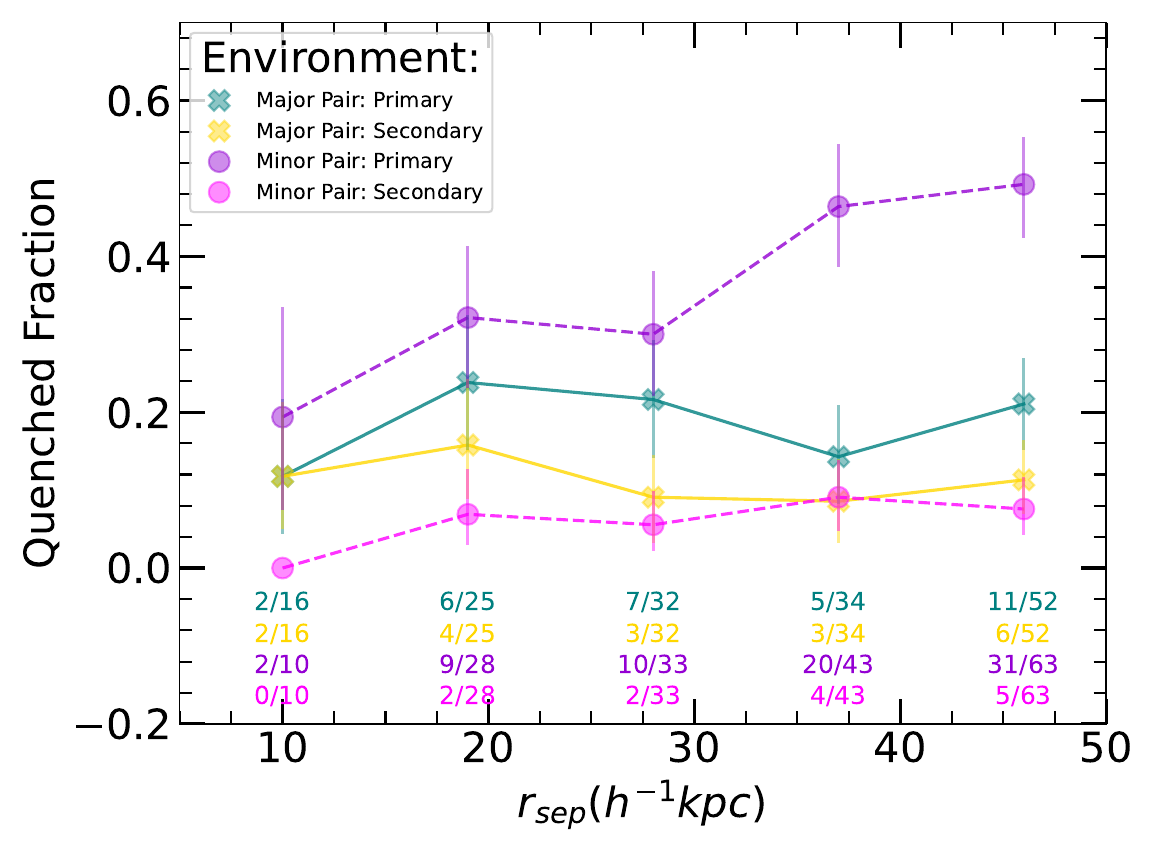}
\caption{The quenched fraction vs. projected separation of close pairs. We separate the close pairs into major primaries (teal crosses), major secondaries (yellow crosses), minor primaries (purple dots) and minor secondaries (magenta dots). Each distribution has been segmented into projected separation bins with a width of 9 $h^{-1} kpc$. The number of suppressed and total number of galaxies in each bin are displayed in the lower region of the figure. Errors have been computed using bootstrap resampling within each bin. There is a general trend of decreasing suppressed fraction as the projected separation decreases amongst the galaxy pairs.}
\label{SF_sep_Major_Minor}
\end{figure}

Following previous works \citep[e.g.][]{Robotham2014, Davies15, Li23}, we classify our close pairs into minor pairs, which have a stellar mass ratio less than 1:4 ($M_{sec}/M_{prim} < 0.25$), and major pairs, which have a stellar mass ratio greater than 1:4 ($M_{sec}/M_{prim} \geq 0.25$). $M_{sec}$ refers to the secondary galaxy in the pair which is the less massive galaxy and $M_{prim}$ refers to the primary galaxy in the pair, the most massive. Minor pairs provide us insight as to the effects of what happens when a more massive galaxy interacts with a comparatively lower-massed galaxy, while major pairs offer insight into the effects of two relatively equally massed galaxies interacting with each other. The distribution of mass ratios among the close pairs is illustrated in Figure \ref{mratio}, where a dashed line delineates the distinction between major and minor pairs. This results in the sample being comprised of $53\%$ as minor pairs, while $47\%$ are classified as major pairs.

By categorising our close pairs into major and minor pairs and identifying primary and secondary galaxies, we can analyse the median distribution of $\Delta sSFR$ while controlling for stellar mass across a range of interaction scales. As depicted in Figure \ref{Delta_Pairs}, in line with the analysis in Sections \ref{subsec:SF_Local} and \ref{subsec:SF_Group}, we examine the variations within the SF population of these galaxies.

The median distributions of $\Delta sSFR$ for major primaries and major secondaries demonstrate notably similar behaviour, as expected. Given their comparable masses, it is anticipated that these two galaxy populations would behave similarly during their interactions. The SF population of major pairs exhibit enhancements in their SF efficiency compared to the field at stellar masses lower than $10^{10} M_{\odot}$, aligning with the findings of \cite{Woods10}, who observed a tendency for major pairs in SDSS to display heightened levels of SF compared to the field. However, at $M_{\star} > 10^{10} M_{\odot}$, the major pairs follow the SF distribution observed in the field.

Conversely, minor pairs exhibit a diverse range of star-forming behaviours. Minor primaries align with the field distribution until $M_{\star} > 10^{10.5} M_{\odot}$, where a decrease in SF efficiency relative to the field is observed. Notably, SF is detectable at larger stellar masses for minor primaries compared to the field in Figure \ref{Delta_Local}, extending our analysis close to a stellar mass of $10^{11.3} M_{\odot}$. This suggests that, although less efficient at these stellar masses, the star-forming population in minor primaries remains active rather than transitioning to a quenched state with no ongoing SF, as observed in the field, or already quenched galaxies undergo ``re-animation'' and see a resurgence in their SF. If SF were quenched at these large stellar masses, it would manifest as a SFR UL value, which is notably lacking in the major primaries relative to the field. This discrepancy underscores the resilience of SF processes within minor primaries, possibly facilitated by mechanisms such as tidal stripping of gas from the less massive member, which may serve to fuel new SF processes \citep{Mihos04, Spilker22}, or more likely, tidal torques help compress the gas reservoirs, fueling new SF, particularly in central regions \citep{Mihos94, Moreno21, Li23}.

Our expanded sample extends to lower stellar masses, where the median distribution demonstrates large SF efficiencies relative to the field. Conversely, at stellar masses greater than $10^{10} M_{\odot}$, minor secondaries exhibit a transition to a less efficient state of SF compared to the field. In contrast to \cite{Davies15}, who noted heightened levels of suppressed SF in minor secondaries when looking at a mass range between $10^{9.5}-10^{11} \: M_{\odot}$, we only detect this at within our largest stellar mass bin. \cite{Davies15} suggests that the SF suppression in minor secondaries occurs in very short ($\lesssim 100 Myr$) and is not evident in long-duration SFR tracers such as the mid-IR, however, it's worth noting that minor secondaries show a relative increase in the amount of UL values compared to other pair populations, suggestive of a diverse range of SF behaviours; encompassing both enhanced and highly inefficient SF processes.

To examine the correlation between proximity and induced SF, we analyse the relationship between the quenched fraction of close pairs and their projected separation ($r_{\rm sep}$). Figure \ref{SF_sep_Major_Minor} illustrates this by separating close pairs into major/minor and primary/secondary classifications, testing the effects of varying interaction scales. This approach is consistent with prior studies that investigate the impact of pair dynamics on SF \citep{Lambas03, Ellison08, Li08, Scudder12, Patton13, Bustamante20, Patton20, Steffen21}. While we acknowledge that stellar mass plays a critical role in quenching, as demonstrated earlier, the sample size was insufficient to separate the data into mass bins, as done by \cite{Davies15}. A larger dataset would enable a more detailed, mass-segregated analysis.

We observe a correlation between projected separation and the quenched fraction, where the quenched fraction of minor pairs decreases as the projected separation decreases. This effect is more pronounced for minor primaries than minor secondaries. Conversely, major pairs do not exhibit any clear trends over the range of projected separations chosen in this analysis. These findings for our minor pairs align with previous studies by \cite{Lambas03, Ellison10, Scudder12, Davies15, Steffen21, Shah2022}, which demonstrated increased levels of star formation as a function of decreasing projected separation, particularly for pairs with separations of $r_{\rm sep} < 30 \: \text{h}^{-1} \text{kpc}$.

We emphasise the analysis of close galaxy pairs is a difficult challenge, as there are numerous factors influencing their evolutionary trajectories, particularly across different stellar mass regimes. A more extensive dataset is necessary to unravel the underlying mechanisms driving these complexities. Moreover, more detailed observational studies on individual interacting close pairs such as detailed HI studies are essential to understanding the observed trends we find, particularly those observed to deviate from the ``typical'' evolutionary pathways.

\section{Discussion}
\label{sec:Discussion}

Our analysis has revealed clear trends in the relationship between environmental factors and galaxy evolution within the local universe. Using a mass-complete sample of galaxies to $z < 0.1$, we have identified differing patterns in SF and quenching processes across various environments.

We observed an increase in SF quenching with higher group membership, indicating that galaxies in denser environments are more likely to have their SF processes suppressed. Specifically, the fraction of quenched galaxies rises with group membership when controlling for stellar mass, highlighting the impact of group environments on SF quenching.

Within the star-forming population, we found variations in specific star formation rates ($\Delta sSFR$) that correlate with group membership. These variations suggest that the group environment significantly influences SF efficiency, with more pronounced effects as group membership increases.

These findings underscore the significant role of group environments in shaping galaxy evolution. The following sections will build on our analysis, discussing the impacts of environment on SF quenching (Section \ref{subsec:Discussion EQE}) and changes within the SF population (Section \ref{subsec:Discussion SFD}) through a more quantitative approach.

\subsection{Environment Quenching Efficiency}
\label{subsec:Discussion EQE}

In Sections \ref{subsec:SFQ_Local} and \ref{subsec:SFQ_Group} we explored the quenched fractions of galaxies in various local galaxy environments and investigated how this influence differs across varying levels of dark matter halo occupation through galaxy group memberships. To further quantify the efficiency of environment, whilst controlling for stellar mass, we make use of the \textit{environmental quenching efficiency} metric, $\epsilon_{\rm env}$. This quantity was initially introduced in \cite{Peng10} and has been utilised in more recent works \cite[e.g.][]{Contini20,Delgado22,Shi2024} to quantify the relative role of environment in quenching galaxies. We define the \textit{environmental quenching efficiency} as:

\begin{equation}
\centering
\label{EQE}
\epsilon_{\rm env} \: (F, SSE, M_{\star}) = \frac{f_{q} (SSE, M_{\star}) - f_{q} (F, M_{\star})}{1 - f_{q} (F, M_{\star})},
\end{equation}

\noindent where $f_{q} (SSE, M_{\star})$ is the fraction of quenched galaxies with stellar mass $M_{\star}$ in a given small-scale environment (SSE), and $f_{q} (F, M_{\star})$ is the same fraction for quenched galaxies in the field (F). Unlike the quenched fraction, which only allows for qualitative comparisons between environments, $\epsilon_{\rm env}$ quantifies the excess of quenched galaxies relative to the field environment, allowing for a more direct comparison 

\begin{figure}[!htb]
\centering
\includegraphics[width=\linewidth]{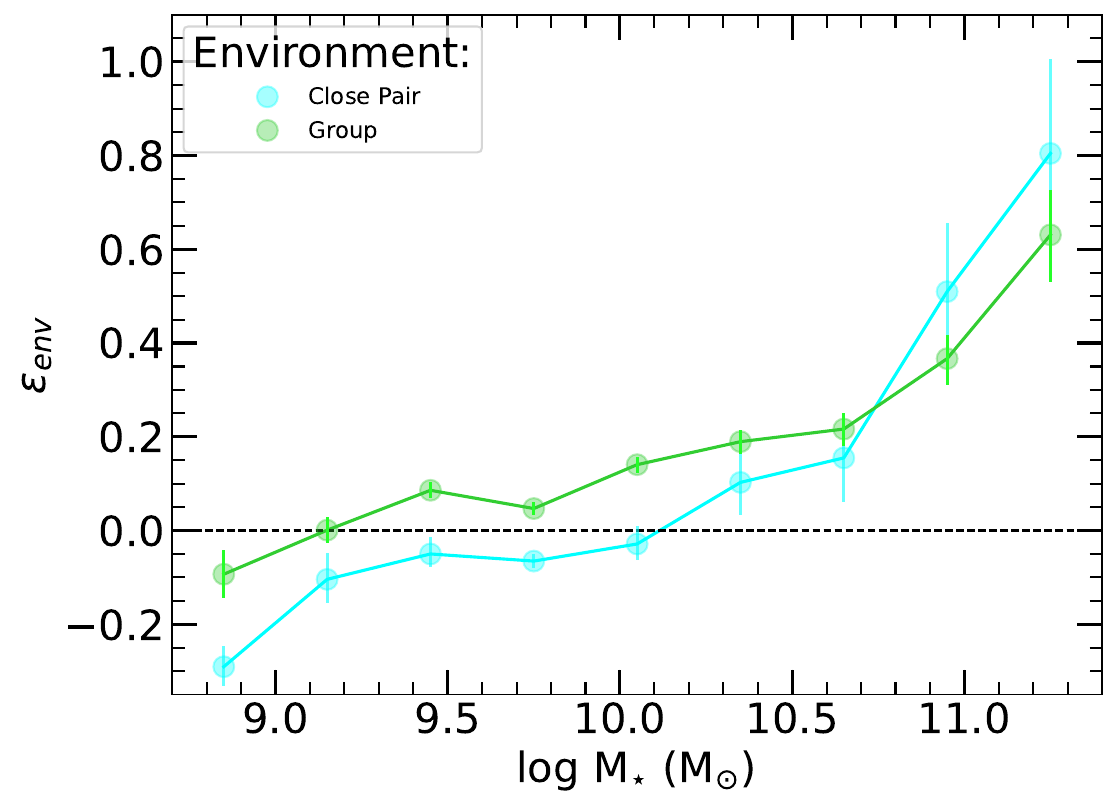}
\caption{The environmental quenching efficiency as a function of stellar mass for close pairs and groups. The environmental quenching efficiency quantifies the fractional difference in the fraction of quiescent galaxies between the field and environmental populations within each stellar mass bin. A positive environmental quenching efficiency indicates a relative increase in the fraction of quiescent galaxies compared to the field, while a negative value suggests a decrease. The black dashed line indicates no difference between the field population. Errors are calculated from the errors of the medians in Figure \ref{Local_QF}.}
\label{Local_EQE}
\end{figure}

\noindent and attributing this excess to physical processes linked to the specific environment. Errors for the $\epsilon_{\rm env}$ are derived from the bootstrapped resampled errors from the relevant quenched fraction results.

Figure \ref{Local_EQE} presents our analysis of the $\epsilon_{\rm env}$ across varying stellar masses for the combined close pair and group samples. Close pairs exhibit a 29\% lower fraction of quenched galaxies in the lowest mass bin compared to the field. This trend persists, with pairs remaining less quenched than the field up to $M_{\star} \sim~10^{10}~M_{\odot}$, after which there is a sharp increase, resulting in an 80\% higher fraction of quenched galaxies at the largest mass bin of $M_{\star} = 10^{11.25}~M_{\odot}$. Conversely, the combined group galaxy sample shows a lower fraction of quiescent galaxies (9\% less) than the field only in the lowest mass bin. The grouped galaxies exhibit increasing $\epsilon_{\rm env}$ values as stellar mass increases, culminating in a 63\% increase in quiescent galaxies in the largest mass bin. Our results of the combined group sample Figure \ref{Local_EQE} match the results of \cite{Delgado22} remarkably well when comparing the same stellar mass range, considering they used a more extended redshift range ($z < 1$). We show a similar small bump in the $\epsilon_{\rm env}$ at $M_{\star} \sim 10^{9.5}~M_{\odot}$, followed by a constant increase to $\epsilon_{env} \sim 0.6$ at $M_{\star} = 10^{11.25}~M_{\odot}$.

\begin{figure}[!hbt]
\centering
\includegraphics[width=\linewidth]{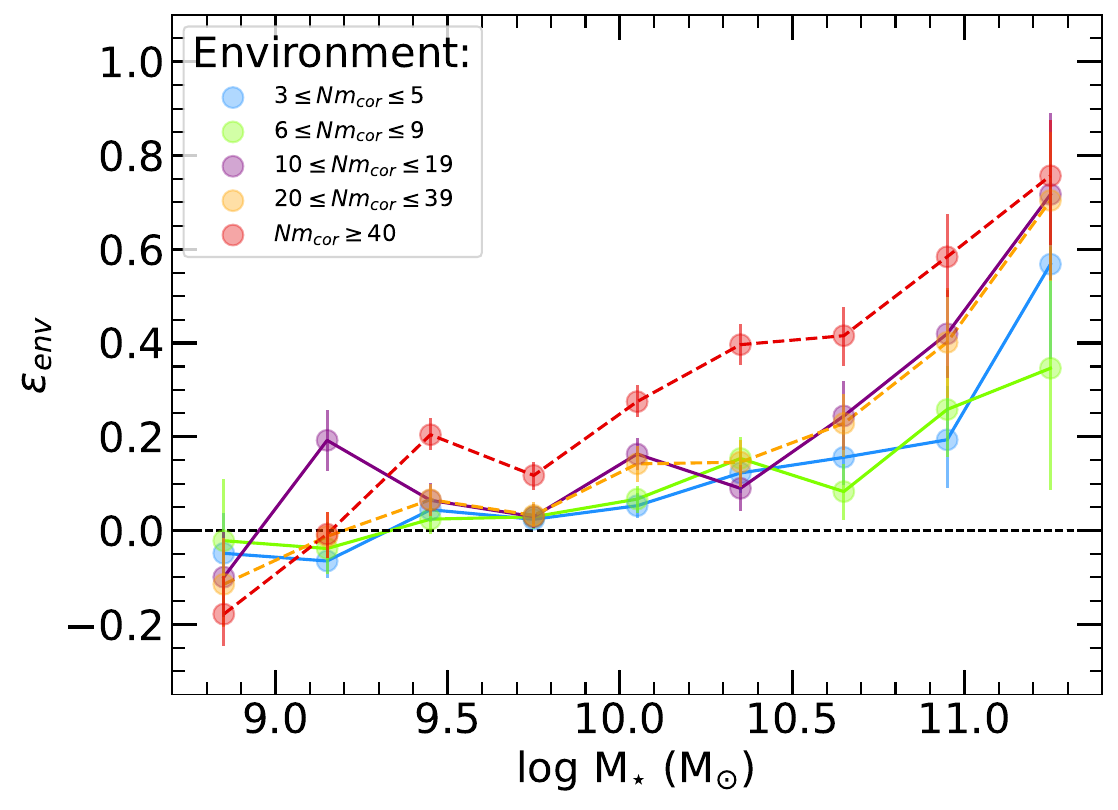}
\caption{The environmental quenching efficiency as a function of stellar mass for the different galaxy group population distributions. The environmental quenching efficiency quantifies the fractional difference in the fraction of quiescent galaxies between the field and differing group populations within each stellar mass bin. A positive environmental quenching efficiency indicates a relative increase in the fraction of quiescent galaxies compared to the field, while a negative value suggests a decrease. The black dashed line indicates no difference between the field population. Errors are calculated from the errors of the medians in Figure \ref{Group_QF}.}
\label{Group_EQE}
\end{figure}

Figure \ref{Group_EQE} separates the combined galaxy groups into their group membership/HOD bins to study  $\epsilon_{\rm env}$ spanning a range of low-density to high-density environments. The left-hand panel shows the distribution for the smaller membership corrected groups, while the right-hand panel shows the larger membership corrected groups. All galaxy groups exhibit a decrease in $\epsilon_{\rm env}$ within the lowest mass bin, typically becoming more pronounced as group membership increases. This is evidenced by a reduction in the excess of quenched galaxies, ranging from 2\% for the 6-9 membership groups to 18\% for the 40+ membership groups, indicating that the physical processes associated with larger group sizes bolster star formation in these low-mass galaxies. As stellar mass increases, there is an observed increase in the excess of quenched galaxies relative to the field. This excess further amplifies with increasing group membership, rising from 35\% for the 6-9 membership groups to 76\% for the 40+ membership groups.

This work extends previous studies on environmental quenching efficiencies, which typically analysed this phenomenon as a function of redshift \cite[e.g.][]{Pinto19,Chartab20}, or a combination of redshift and overdensity \cite[e.g.][]{Peng10,Quadri12,Kovac14,Darvish16,Kawinwanichakij17,Shi2024}. Most of these works do not study the very nearby universe and their redshift samples are mostly $z > 0.5$. Typically, studies examining $\epsilon_{\rm env}$ as a function of redshift and stellar mass have demonstrated that environmental quenching efficiency is more pronounced at lower redshifts and increases with stellar masses. Similarly, research focusing on $\epsilon_{\rm env}$ as a function of overdensity ($1 + \delta$) has also traced the effects of environment, albeit in a slightly different manner, showed that $\epsilon_{\rm env}$ increases with both overdensity and stellar mass. In this study, we have developed and introduced a comprehensive description of the local universe's environment through the meticulous construction of our galaxy groups. Consequently, the relatively high environmental quenching efficiencies observed at larger stellar masses in our work further emphasize the significant role of environment in the nearby universe, as suggested by prior studies. The size of our sample results in small uncertainties in most stellar mass bins, providing arguably the cleanest and clearest benchmark for aggregate behavior in different group environments currently available.

\begin{figure}[!hbt]
\centering
\includegraphics[width=\linewidth]{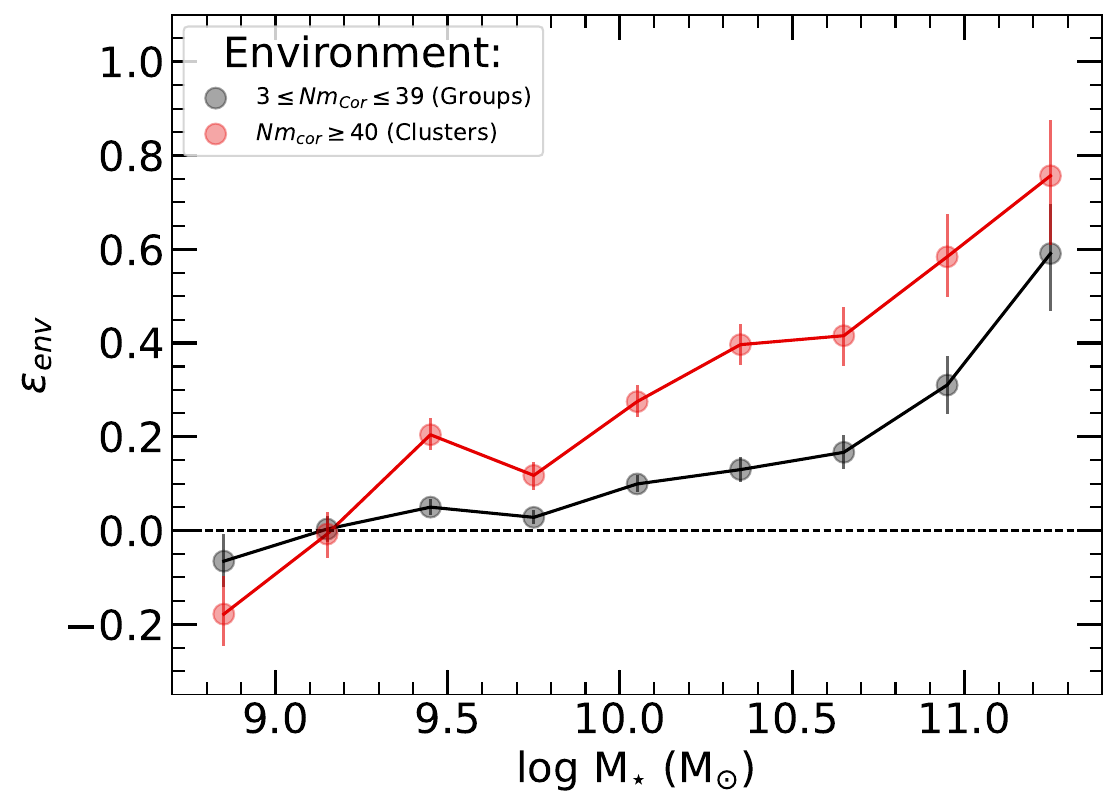}
\caption{The environmental quenching efficiency ($\epsilon_{env}$) as a function of stellar mass for galaxy groups with total corrected members between 3 and 39 and galaxy clusters, having more than 40 corrected members. The environmental quenching efficiency quantifies the fractional difference in the fraction of quiescent galaxies between the field and group/cluster populations within each stellar mass bin. A positive environmental quenching efficiency indicates a relative increase in the fraction of quiescent galaxies compared to the field, while a negative value suggests a decrease. The black dashed line indicates no difference between the field population. Errors are calculated from the errors of the medians in Figure \ref{Group_QF}.}
\label{Group_Cluster_EQE}
\end{figure}

The only prior research to our knowledge that is more closely aligned to the exact works in this study is that of \cite{Balogh16}, \cite{Contini20} and \cite{Delgado22}. These studies define environment using either galaxy groups/clusters, and/or halo masses, which allows for comparison with some of our results. In Figure \ref{Group_Cluster_EQE}, we build upon the works of \cite{Balogh16}, who utilised observational results, and \cite{Contini20}, who used simulations to compare the environmental quenching efficiency of grouped and cluster galaxies at $z=0.9$. For this comparative analysis, we define our 40+ membership groups as cluster galaxies and groups with ($3 \leq Nm_{cor} \leq 39$) as the group galaxy sample, justified by the halo occupation distribution and halo mass relation discussed in Sections \ref{subsec:Halo_Mass} and Appendix \ref{sec:Appendix Halo Mass Var} to best match the halo mass range within these studies.

Compared to the $z=0.9$ results of \cite{Balogh16} and \cite{Contini20}, as shown in Figure \ref{Group_Cluster_EQE} we find similar results within our given mass range, especially for the cluster galaxies. However, our sample group galaxies display a significantly higher quenched excess at larger stellar masses. This can be expected when contrasting $0 < z < 0.1$ to $z=0.9$ due to the evolving environmental dependence in the nearby universe. This leaves room for future comparisons with simulations and other larger observational results of the nearby universe, such as from 4HS \citep{ENTaylor23}. A noteworthy point of difference is observed in the mass range of $10.0 \leq M_{\star} \leq 10.5$, where our groups exhibit a much lower excess of quiescent galaxies. We suspect this point of difference results from the inclusion of lower membership groups, which are not accounted for in \cite{Balogh16} and \cite{Contini20}, as their groups range from $13.5 < \log M_{halo} (M_{\odot}) < 14.0$, which would be more similar to our $10 \leq Nm_{cor} \leq 39$ groups. Consequently, we encapsulate a lower range of halo masses, where these environmental effects may be less pronounced in this mass range.

Overall, we observe an evolving population of quiescent galaxies as a function of environment. While the changing stellar mass function between the field and groups can contribute to the results by populating these environments with more quiescent galaxies \citep{Mehmet15, Etherington17, Papovich18}, the relative number of high-mass galaxies across different group populations controls for this variation. This control is similarly maintained in the low-mass range. Thus, we may be observing the influence of physical processes tied to these environments, where an increased fraction of quiescent galaxies suggests the effects of SF pre-processing. However, as noted by \cite{Calvi13}, the stellar mass function may not significantly vary across different environments, which could further increase the relevancy of our results.

\subsection{Relative Star Formation Deficiency  - The Pre-Processing of Star Formation}
\label{subsec:Discussion SFD}

In Sections \ref{subsec:SF_Local} and \ref{subsec:SF_Group} when comparing the SF group galaxies to those in the field, we observe a trend towards decreased SF efficiency as a function of environment across most stellar masses. This is evident from the relative decrease in $\Delta sSFR$ observed in Figures \ref{Delta_Local} and \ref{Delta_Group} for the star-forming population. To quantitatively assess this shift in star-forming behaviour across different environments, we introduce a new metric, the \textit{star formation deficiency} ($\epsilon_{SFD}$). This metric follows the same motivations as the \textit{environmental quenching efficiency}, but rather than focusing on the excess of quenched galaxies relative to the field, it focuses on the relative difference in aggregate star-forming behaviour relative to the field. We define the \textit{star formation deficiency} as:

\begin{equation}
\centering
\label{SFD}
\epsilon_{SFD} \: (F, SSE, M_{\star}) = \frac{\widetilde{\Delta sSFR} (F, M_{\star}) - \widetilde{\Delta sSFR} (SSE, M_{\star})}{1 - \widetilde{\Delta sSFR} (F, M_{\star})},
\end{equation}

\noindent where $\widetilde{\Delta sSFR} (SSE, M_{\star})$ is the median $\Delta sSFR$ of galaxies with stellar mass $M_{\star}$ in a given small-scale environment (SSE), and $\widetilde{\Delta sSFR} (F, M_{\star})$ is the same median but for $\Delta sSFR$ in the field (F). Positive values of $\epsilon_{SFD}$ represent a fractional excess of decreased SF efficiency in the small-scale environments relative to the field, while negative values indicate a fractional excess of increased SF efficiency. \textit{Thus, when the star formation deficiency metric is negative, it reflects an enhancement of star formation efficiency compared to the field}. The star formation deficiency is intended to gauge the extent to which environmental factors influence SF efficiency within the star-forming galaxy population across various environmental settings while controlling for stellar mass. \textit{This metric provides valuable insights into the impact of environmental processes, such as pre-processing, on SF dynamics}, as without pre-processing occurring in local environments, we would not expect to observe a global change in the star-forming population, and thus the composition of the star-forming population would remain relatively constant, even across the likely varying stellar mass functions of the different environments. Errors for the $\epsilon_{\rm SFD}$ are derived from the bootstrapped resampled errors from the relevant $\Delta sSFR$ results.

\begin{figure}[!hbt]
\centering
\includegraphics[width=\linewidth]{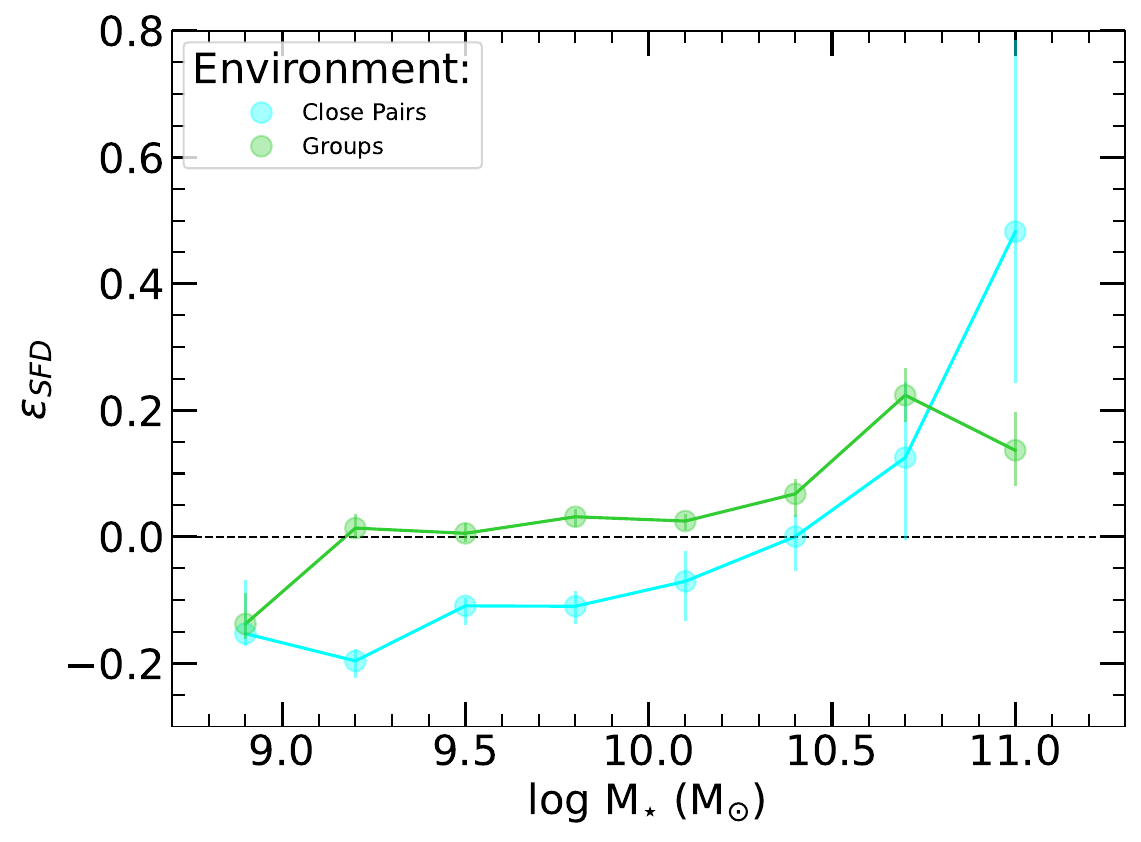}
\caption{The star formation deficiency as a function of stellar mass for close pairs and groups. The star formation deficiency quantifies the fractional difference in median sSFR behaviour between the field and environmental populations within each stellar mass bin. A positive star formation deficiency indicates a relative decrease in star formation activity compared to the field, while a negative value suggests an increase. The black dashed line indicates no difference between the field population. Errors are calculated from the errors of the medians in Figure \ref{Delta_Local}.}
\label{Local_SFD}
\end{figure}

Figure \ref{Local_SFD} presents our analysis of the $\epsilon_{SFD}$ across varying stellar masses for the combined close pair and group samples. We observe that the close pairs exhibit SF enhancements of 15\% relative to the field in the low mass regime. This enhancement gradually diminishes to zero at $\log M_{\star} = 10.4 \: M_{\odot}$, after which the pairs experience a rapid increase in \textit{star formation deficiency}, becoming increasingly deficient compared to the field, with the deficiency peaking at 48\%, albeit with a larger uncertainty. In the lowest stellar mass bin, the group sample shows a similar enhancement in star formation as the close pairs, with a 14\% increase. From the next mass bin onward, we observe a sharp rise in the \textit{star formation deficiency}, moving above the field line and peaking at a 22\% relative deficiency in higher mass galaxies.

\begin{figure}[!hbt]
\centering
\includegraphics[width=\linewidth]{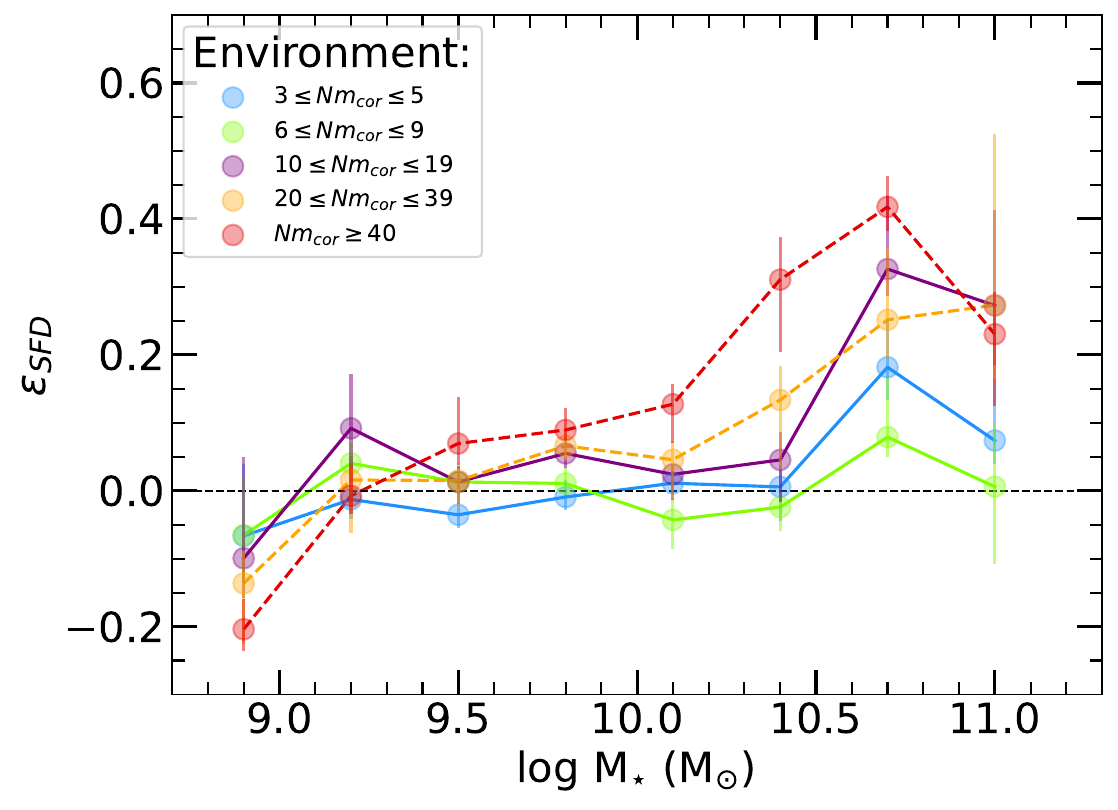}
\caption{The star formation deficiency ($\epsilon_{SFD}$) as a function of stellar mass for the varying group environments. The star formation deficiency quantifies the fractional difference in median sSFR behaviour between the field and group populations within each stellar mass bin. A positive star formation deficiency indicates a relative decrease in star formation activity compared to the field, while a negative value suggests an increase. The black dashed line indicates no difference between the field population. Errors are calculated from the errors of the medians in Figure \ref{Delta_Group}. The overall behaviour of the different group galaxies remains constant but increases in their relative effects as a function of group membership.}
\label{Group_SFD}
\end{figure}

Figure \ref{Group_SFD} further breaks down the observed effects on the group galaxies by separating out into the group membership/HOD bins. The grouped galaxies typically exhibit larger relative $\epsilon_{SFD}$ as group membership increases, ranging from a maximum star formation deficiency of only 8\% for the 6-9 membership groups to 42\% for the 40+ membership groups. Additionally, we observe that the increase in $\epsilon_{SFD}$ manifests at lower stellar masses as group membership increases. 

We also observe increases in the relative SF efficiency observed in each of the lowest mass bins, ranging from 7\% for the 6-9 membership groups to 20\% for the 40+ membership groups. Notably, the 6-9 membership groups display the smallest relative changes in their star-forming behaviour, not the 3-5 membership groups as one might expect, a phenomenon warranting further exploration in subsequent studies.

The environments inhabited by both pairs and groups; particularly those characterised by low membership, serve as valuable indicators of SF pre-processing as these are less impacted by the possible effects of changing stellar mass functions. Traditionally, changes in the star-forming population relative to the field are not associated with such low environmental densities, therefore, the observation of these subtle yet statistically significant deviations from the field, which intensify with increasing environmental density, suggests the potential tracing of galaxies' star-forming activity from their environment. This hypothesis is consistent with previous studies \cite[e.g.][]{Cortese06,Oman2016,Bakels21}, which propose that galaxies do not undergo immediate quenching upon entering highly dense environments such as clusters. Instead, their SF undergoes gradual pre-processing in smaller group environments, with these effects becoming significantly amplified upon their transition into denser environments.

As previously mentioned, the smallest and largest mass bins exhibit the lowest statistical significance. Despite our meticulous approach to handling these masses, there remains inherent uncertainty in the results. These particular mass ranges likely require a larger sample size to enhance confidence in our findings. Upcoming surveys like the 4-MOST Hemisphere Survey (4HS), as noted by \cite{ENTaylor23}, could address this need, offering greater robustness to our conclusions across all stellar masses.

\section{Summary \& Conclusions}
\label{sec:Sum_Con}

In this study, we delved into the intricate interplay between galaxies and their local environment in the $z < 0.1$ universe. Our analysis included a sample of 24,676 galaxies spread across a 384 degree square region, centred on the South Galactic Pole. Drawing data from prominent spectroscopic surveys, primarily that of 2dFGRS and GAMA (G23 Field). The integration of WISE-measured sources, with a cross-match of a 93\% completeness rate, furnished us with robust mid-IR measurements providing us with galaxy properties such as IR colours, stellar masses and star formation rates, all previously derived from \cite{Jarrett2019,Jarrett23} and Cluver et al. (in prep), respectively.

We utilised a FoF python algorithm \citep[{\tt FoFpy};][]{Lambert20}, fine-tuned with inputs from cosmological dark matter simulations running Millennium and SAGE, to delineate galaxy groups. Moreover, our approach accounted for the magnitude limitations of the 2dFGRS and GAMA surveys by utilising simulations to categorise our completeness, allowing us to implement a correction factor to produce galaxy groups with a statistically based correction on the number of members associated with the galaxy groups. This information provides us with a more accurate constraint on the associate group environments, thus rectifying potential biases in the membership count of our galaxy groups. Additionally, we searched for close pairs within our sample, delineating the extreme ramifications of galactic interactions, particularly those occurring within a proximity of $r_{\rm sep}<50$ $\text{h}^{-1}\text{kpc}$ and $v_{\rm sep}< 500 \: \text{km s}^{-1}$. \\

Our primary results are as follows: 

\begin{enumerate}
\item We found 1,413 galaxy groups comprising 8,962 galaxies and employed corrections to the number of members of each group as a function of redshift, the number of observed members and the spectroscopic field in which they reside. A total of 36\% of all galaxies were found to be a part of group environments. 

\item Our close pair search yielded 337 pairs (674 galaxies), this resulted in 15,469 of the galaxies not associated with either a group or close pair environment, making up our field sample.

\item We demonstrated that the fraction of quenched galaxies increases as a function of environment, where more dense environments exhibit larger quenched fractions when controlling for stellar mass. The offsets of the quenched fractions between the field and galaxy groups, as well as between the field and close pairs, increased with increasing stellar mass from $M_{\star} \geq 10^{9.75} M_{\odot}$.

\item We find evidence of star-formation pre-processing when analysing the relative differences in $\Delta sSFR$, which represents the dex difference between the star-forming main sequence fit and the measured sSFR. Our analysis shows an overall decrease in $\Delta sSFR$ within group galaxies compared to isolated field galaxies, particularly at larger stellar masses. The offset between group and field populations increases with galaxy group membership. Additionally, the initial offset in $\Delta sSFR$ between group and field populations occurs at lower stellar masses as galaxy group membership increases.

\item Our analysis of $\Delta sSFR$ in the star-forming population of close pairs reveals diverse SF behaviours when differentiating between major/minor pairs and primary/secondary galaxies. In major pairs, both primary and secondary galaxies exhibit increased star-formation efficiencies, as indicated by an overall increase in $\Delta sSFR$. Conversely, in minor pairs, the primary galaxy shows signs of star-formation inefficiency, while the secondary galaxy demonstrates increased star-formation efficiency.

\item In Figure \ref{EQE} we employ the environmental quenching efficiency introduced in \cite{Peng10} to quantify the excess of quiescent galaxies in different environments compared to the field. Close pairs exhibit up to 29\% fewer quiescent galaxies at the lowest stellar masses bin, but this trend reverses at higher masses, with up to 80\% more quiescent galaxies. Low membership groups show minimal environmental quenching efficiency at low masses. However, at a stellar mass of $10^{10} M_{\odot}$, these populations deviate from the field. Similar to the pairs, the larger membership groups display slight decreases in the excess of quiescent galaxies (up to 18\%) at the lowest mass bin, but the excess if quiescent galaxies rapidly increases rapidly as mass increases, reaching up to 76\%.

\item In Section \ref{subsec:Discussion SFD}, we introduced a new metric to quantify the relative impact on the star-forming population of galaxies relative to the environment and test for pre-processing of star formation. The \textit{star formation deficiency} ($\epsilon_{SFD}$) metric indicates the disparity between the median $\Delta sSFR$ of various environments and the field population. Our findings shown in Figure \ref{Local_SFD} indicate a range of trends for pairs, from a median increase in star formation of 15\% relative to the field at low stellar masses to a decrease of 48\% at larger stellar masses. In Figure \ref{Group_SFD} the relative effects on group galaxies heightened with increasing membership, with a relative increase in star formation of 7\% at the lowest stellar mass bin within the smaller membership groups, increasing to 20\% for the largest membership groups at low stellar masses. At larger stellar masses, we observed a decrease in star formation ranging from 8\% for the small groups to 42\% for the largest groups, quantifying the average change in star-forming behaviour across different environments whilst controlling for stellar mass.

\end{enumerate} 

Despite the robust determination of environmental metrics and highly complete redshift information, the primary limitation of this study lies in its statistical fidelity. While we possess a substantial dataset for intermediate stellar masses ($10^{9.5} M_{\odot}$ $<$ $M_{\star}$ $<$ $10^{10.75} M_{\odot}$), this isn't equally true for lower and higher stellar mass ranges, where our statistical power falls short of providing statistically significant insights into galactic evolution pathways across cosmic epochs. As alluded to, future spectroscopic surveys such as 4-MOST Hemisphere Survey \cite{ENTaylor23} promise to address this shortfall by offering heightened completeness across substantially larger areas, thereby furnishing us with the requisite statistical robustness to fundamentally quantify the effects of environment in the local universe, and the addition of neutral gas analysis through the future use of SKA H{\tt I} and its precursors MeerKAT and ASKAP, will further establish the distinct ways in which galaxies are influenced by their surroundings.

This research underscores the significant impact of environmental factors on galaxy evolution. Our findings reveal that, even when stellar masses are held constant, the local environment correlates with changes in star-forming properties, illustrating that galaxies do not evolve independently over cosmic time but rather evolve through a complex interplay between internal dynamics and external mechanisms.

\section{Acknowledgements}
\label{sec:Acknowledgements}

We gratefully acknowledge the significant contributions and invaluable insights of Thomas Jarrett over the past few decades in infrared astronomy. His legacy will continue to inspire and live on throughout our works and within in our memories.

We thank the anonymous referee for helpful comments and suggestions that have improved the content and clarity of this paper. M.E.C. is a recipient of an Australian Research Council Future Fellowship (project No. FT170100273) funded by the Australian Government. T.H.J. acknowledges support from the National Research Foundation (South Africa). D.J.C. is a recipient of an Australian Research Council Future Fellowship (project No. FT220100841) funded by the Australian Government. This publication makes use of data products from the Wide-field Infrared Survey Explorer, which is a joint project of the University of California, Los Angeles, and the Jet Propulsion Laboratory/California Institute of Technology, funded by the National Aeronautics and Space Administration. GAMA is a joint European-Australasian project based around a spectroscopic campaign using the Anglo-Australian Telescope. The GAMA input catalogue is based on data taken from the Sloan Digital Sky Survey and the UKIRT Infrared Deep Sky Survey. Complementary imaging of the GAMA regions is being obtained by a number of independent survey programmes including GALEX MIS, VST KiDS, VISTA VIKING, WISE, Herschel-ATLAS, GMRT and ASKAP providing UV to radio coverage. GAMA is funded by the STFC (UK), the ARC (Australia), the AAO, and the participating institutions. The GAMA website is \url{https://www.gama-survey.org/}. Based on observations made with ESO Telescopes at the La Silla Paranal Observatory under programme ID 177.A-3016. This research has made use of {\tt python} (\url{https://www.python.org}) and python packages: {\tt astropy} \citep{Astropy13,Astropy18}, {\tt matplotlib} \url{http://matplotlib.org/} \citep{Hunter07}, {\tt NumPy} \url{http://www.numpy.org/} \citep{Walt11}, and {\tt SciPy} \url{https://www.scipy.org/} \citep{Virtanen20}.

\bibliographystyle{apj}
\bibliography{ref}

\appendix
\section{\texorpdfstring{$\alpha$} E Equations}
\label{sec:Appendix alpha}

\renewcommand{\thetable}{A\arabic{table}}
\renewcommand{\thefigure}{A\arabic{figure}}

\begin{table}[!hbt]
    \centering
    \caption{Table of 2dFGRS and GAMA G23 group completeness scale ($\alpha$) equations for various group sizes found by the FoF method, where $a$, $b$, and $c$ are the coefficients in the completeness scale equations: $\alpha = az^{2} + bz^{2} + c$.}
    \label{Group Completeness Equations}
    {\tablefont\begin{tabularx}{\columnwidth}{CCCC}
        \hline \hline
        \multicolumn{4}{c}{\textbf{2dFGRS}} \\
        \hline
        No. of Members & a & b & c \\
        \hline
        3--5 & 185.51 & -5.60 & 1.02 \\
        6--9 & 230.89 & -3.32 & 1.01 \\
        10--19 & 268.23 & -4.90 & 1.02 \\
        20--39 & 321.52 & -7.43 & 1.04 \\
        40+ & 261.74 & -2.63 & 0.98 \\
        \hline
        \multicolumn{4}{c}{\textbf{GAMA G23}} \\
        \hline
        No. of Members & a & b & c \\
        \hline
        3--5 & 39.82 & 0.79 & 0.97 \\
        6--9 & 89.63 & -1.11 & 0.98 \\
        10--19 & 83.11 & 0.28 & 0.97 \\
        20--39 & 87.21 & 0.22 & 0.98 \\
        40+ & 87.61 & 0.18 & 0.98 \\
        \hline \hline
    \end{tabularx}}
\end{table}

\renewcommand{\thetable}{B\arabic{table}}
\renewcommand{\thefigure}{B\arabic{figure}}

\begin{figure*}[!hbt]
\centering
\includegraphics[width=\linewidth]{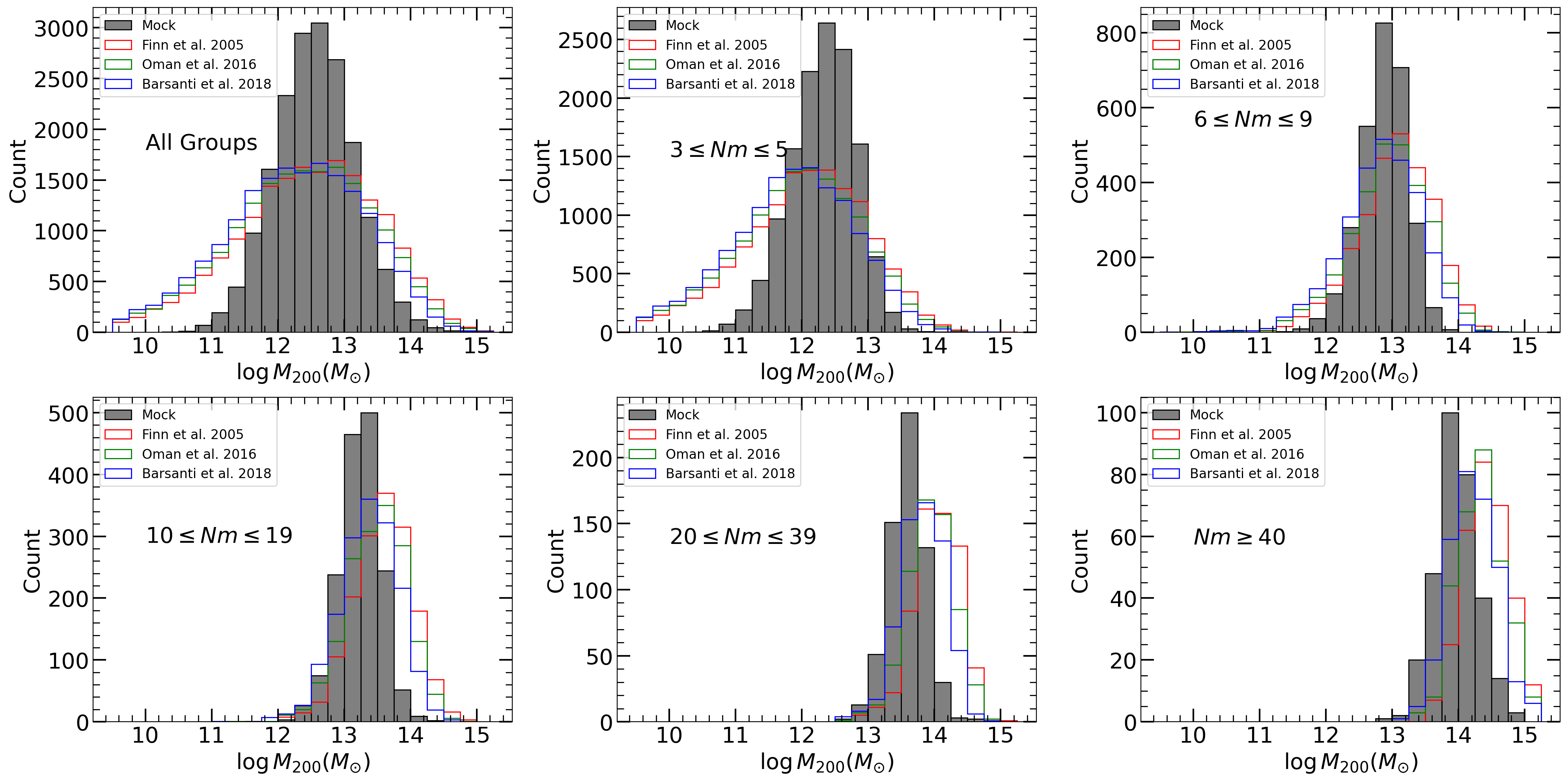}
\caption{The distribution of halo masses derived from a combined set of 25 mock observations. The histograms are categorised into the same galaxy group membership bins utilised throughout this study. The grey distribution represents halo masses obtained from the mock observations, while the other distributions depict halo masses derived from experimental techniques applied to the same galaxy groups.}
\label{M200 Hist}
\end{figure*}

\begin{figure*}[!hbt]
\centering
\includegraphics[width=\linewidth]{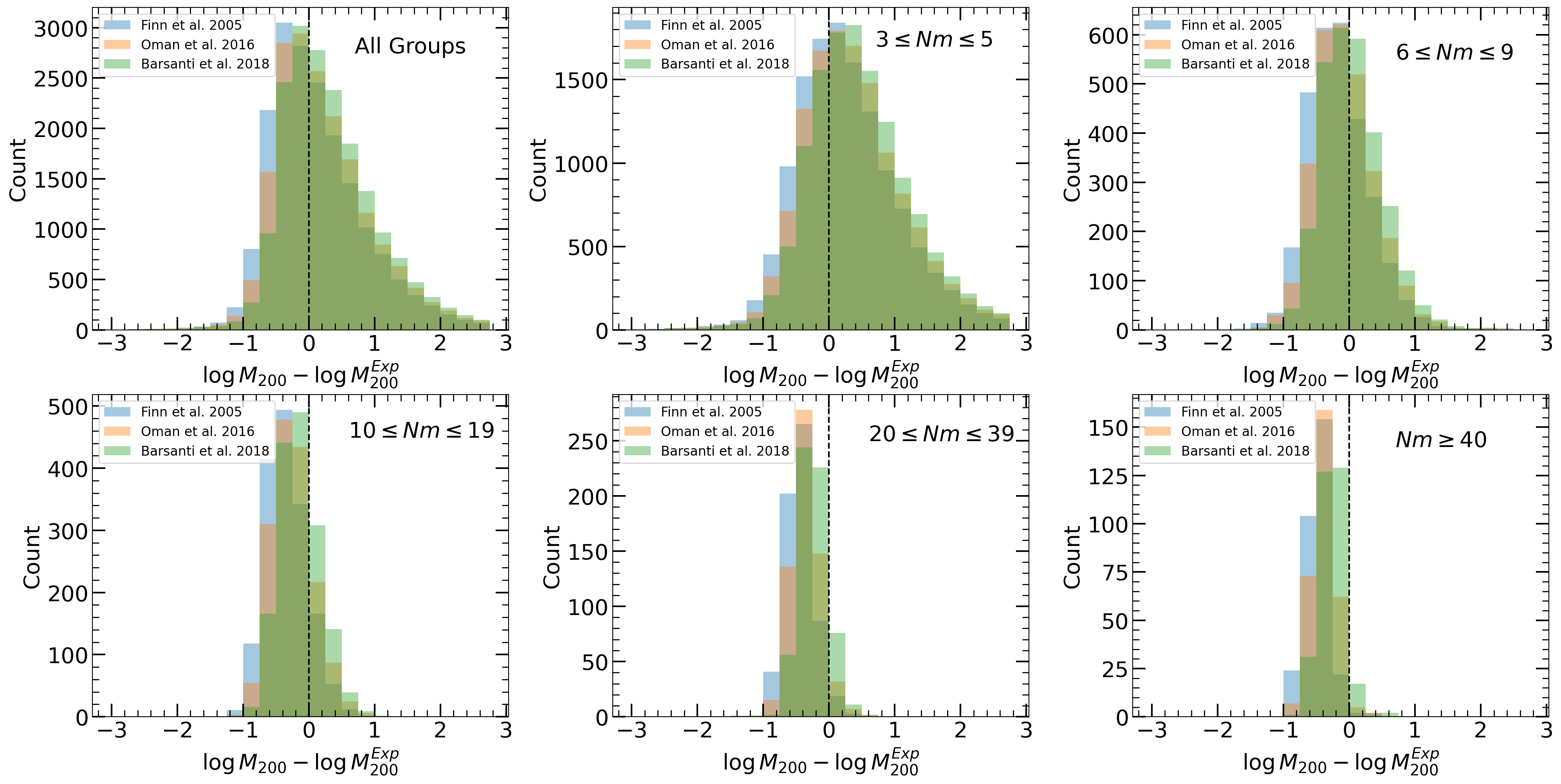}
\caption{The distribution of the dex difference between the mock halo masses and those obtained from various experimental techniques for the combined set of 25 mock observations. The histograms are categorised into the same galaxy group membership bins utilised throughout this study. The different methods show similar results, having noticeable variance in low membership groups, particularly in underestimating halo masses. As membership increases, variance decreases, but a trend towards overestimating halo masses becomes apparent.}
\label{Delta M200}
\end{figure*}

As outlined in Section \ref{subsec:Group Completeness}, we have developed equations for $\alpha$, the group multiplicity correction factor. The $\alpha$ parameter aims at correcting the observed number of galaxies within a group to a more statistically accurate representation of the number of members that have $M_{\star} > 10^{8} \: M_{\odot}$. Table \ref{Group Completeness Equations} presents the coefficients of the 2$^{nd}$ order polynomial used to derive the correction factor for a given amount of detected galaxies across redshift space.

\section{Variance in Halo Mass}
\label{sec:Appendix Halo Mass Var}

\renewcommand{\thetable}{C\arabic{table}}
\renewcommand{\thefigure}{C\arabic{figure}}

\begin{figure*}[!htb]
\centering
    \begin{minipage}[t]{0.5\textwidth}
        \centering
        \includegraphics[width=\textwidth]{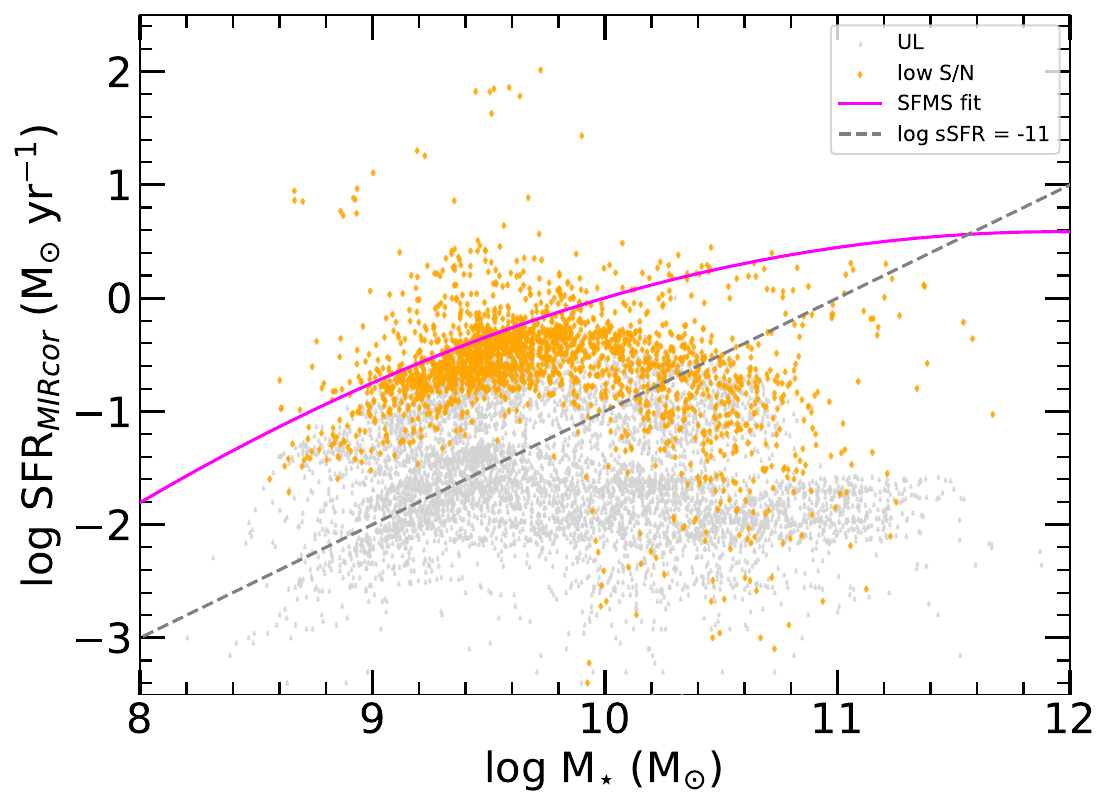}
    \end{minipage}\hfill
    \begin{minipage}[t]{0.5\textwidth}
        \centering
        \includegraphics[width=\textwidth]{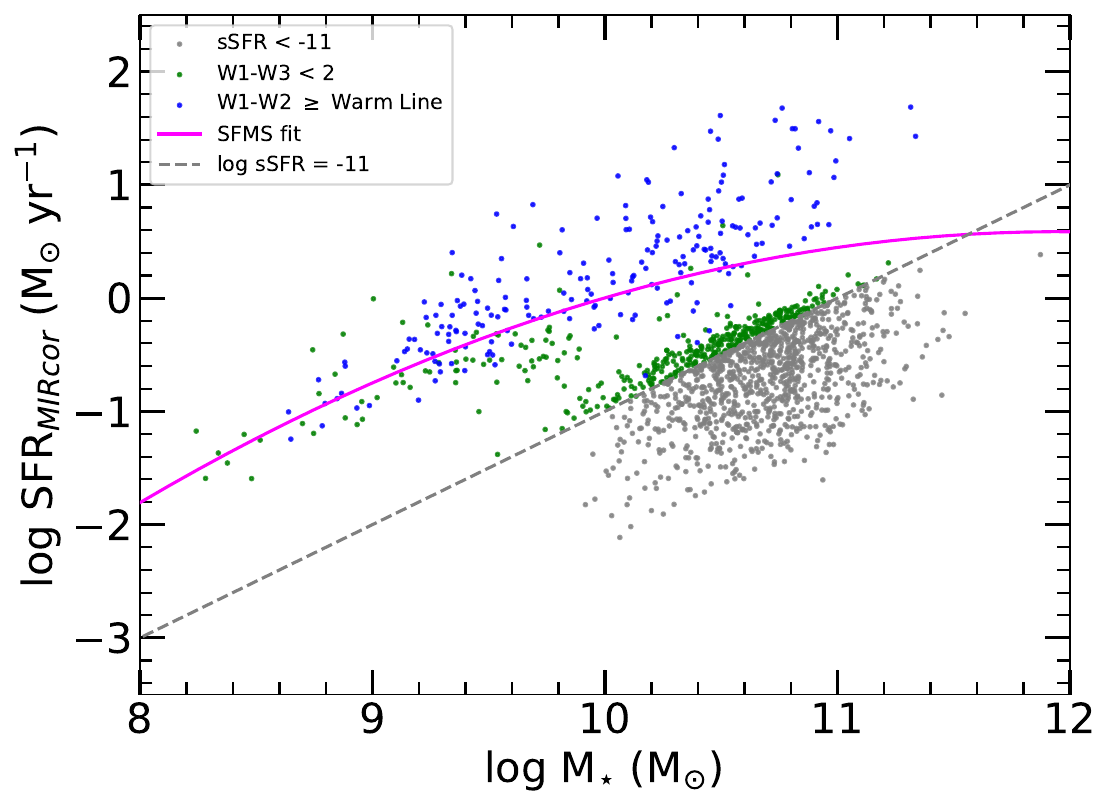}
    \end{minipage}
\caption{This distribution of galaxies that were removed when forming the distribution of the SFMS. The left-hand panel shows the galaxies removed by the low-quality measurement removal. This includes the low S/N (S/N $<$ 2) sources as orange points and SFR UL as grey downward arrows. The right panel shows those removed due to the various relations delineating the SFMS. The grey points indicate those with a SFR less than the $\log sSFR = -11$ line, the blue points are galaxies that are above the colour-colour warm line in Equation \ref{W1W2_W1W3_Equation} and the green points indicate stellar-dominated galaxies (W1-W3 $\leq$ 2). The magenta and dashed magenta lines indicate the SFMS polynomial fit and the 2$\sigma$ offset.}
\label{SFMS_Appendix}
\end{figure*}

The mean values of halo masses, as presented in Table \ref{Halo Mass - Members}, are derived from the distribution depicted in Figure \ref{M200 Hist}. This analysis contrasts the halo mass estimates obtained from theoretical models with those derived from experimental methodologies outlined in Equations 10 in \cite{Finn05}, 3 in \cite{Oman2016}, and 1 in \cite{Barsanti18}. These comparisons underpin the variations in halo mass determination.

Notably, there is a substantial variance observed in groups between the mock values and those derived from observational methods for the low membership groups, with a tendency to underestimate halo masses. Conversely, groups with 6-9 and 10-19 members demonstrate relatively accurate estimations, possibly indicating that the halo mass-dispersion relations were optimised for groups of these sizes. However, larger groups comprising 20-39 and 40+ members tend to overestimate halo masses, albeit with reduced variance.

Figure \ref{Delta M200} illustrates the disparity between experimental halo mass estimations and those obtained from theoretical models, thus highlighting the aforementioned variances across group sizes. Particularly in low membership groups, we see offsets in the halo mass exceeding 2.5 dex, indicating significant discrepancies. While large membership groups exhibit less variance, they are characterised by systematic offsets. This prompts consideration regarding the adequacy of velocity dispersion as a means of halo mass determination, particularly in small and large membership groups as shown.

The derivation of the different experimental halo mass-dispersion relations stems from derived relations from different theoretical frameworks. The consistent variance observed in halo mass estimates across these methodologies implies that the discrepancy originates from the inherent complexity in the halo mass-dispersion relation, rather than being attributable to an error in the theoretical model used.

Future efforts will focus on refining experimental techniques for determining halo mass, aiming to enhance the accuracy and reliability of these estimations. By integrating other measurements that scale with halo mass (such as group membership, stellar mass, etc.), we will aim to improve the versatility and robustness of observation halo mass estimations.

\section{SFMS Selection Cuts}
\label{sec:Appendix SFMS}

This appendix provides an overview of the distributions of galaxies that were excluded in Section \ref{subsubsec:Star-Forming Main Sequence} to establish the SFMS. Figure \ref{SFMS_Appendix} delves into the distribution of mass-complete galaxies that did not undergo fitting within the SFMS framework. The left-hand panel of the figure showcases galaxies removed from the SFMS due to measurement quality concerns, incorporating low S/N (S/N $<$ 2) measurements depicted as orange diamonds, and UL on Star Formation Rates (SFR) denoted by downward grey arrows. On the other hand, the right-hand panel illustrates galaxies excluded based on two WISE colour criteria: those with W1-W3 $\leq$ 2, portrayed in green, signifying dusty or stellar-dominated galaxies, and those with W1-W2 $\geq$ the threshold defined by Equation \ref{W1W2_W1W3_Equation}, displayed in blue, representing infrared-warm galaxies typically associated with an overestimation of their SFRs. Additionally, the right-hand panel presents distributions plotted in grey, indicative of quenched galaxies, identified by a sSFR $<$ -11. The removal of these selections collectively contributes to the formation of the SFMS.

\section{Upper Limits Handling}
\label{sec:Appendix UL}

\renewcommand{\thetable}{D\arabic{table}}
\renewcommand{\thefigure}{D\arabic{figure}}

The reasons for including the UL values as outlined in \cite{Cluver2020} when analysing the quenched fraction is as follows: SFR value UL typically arise either when a reliable W1-W3 colour is present, but a significant stellar continuum dominates the W3 band, or when there is a W3 flux below the detectable threshold, which usually translates to minimal or no star formation present in the host galaxy. For galaxies with a stellar mass of $\log M_{\star} \geq 10 \: M_{\odot}$, the sensitivity of the W3 band to redshifts below $z < 0.1$ allows for the detection of any galaxy with ongoing star formation within this mass range, hence, any UL in this mass range indicates a quenched galaxy, and thus all SFR UL in this mass range are included in the analysis. For galaxies with a stellar mass of $\log M_{\star} < 10 \: M_{\odot}$, the inclusion of SFR UL as either star-forming or quenched depends on redshift, necessitating an assessment of each source to determine whether the SFR could feasibly be detected given its distance/W3 flux. If detectable, such measurements are considered usable and are included as part of the quenched population. Otherwise, if they fail the test, we conservatively include these galaxies as part of the star-forming population, as we do not know the true position along the SFR-SM plane. 

The SFR UL are given the largest possible SFR value given the detection and could potentially exist anywhere below its currently associated value in the SFR plane. Thus, for this population of galaxies, we exercise caution in handling these sources conservatively, placing those with uncertainty as part of the star-forming population.

\section{Close Pairs In Groups}
\label{sec:Appendix Pairs}

\renewcommand{\thetable}{E\arabic{table}}
\renewcommand{\thefigure}{E\arabic{figure}}

\begin{figure}[!htb]
\centering
\includegraphics[width=\linewidth]{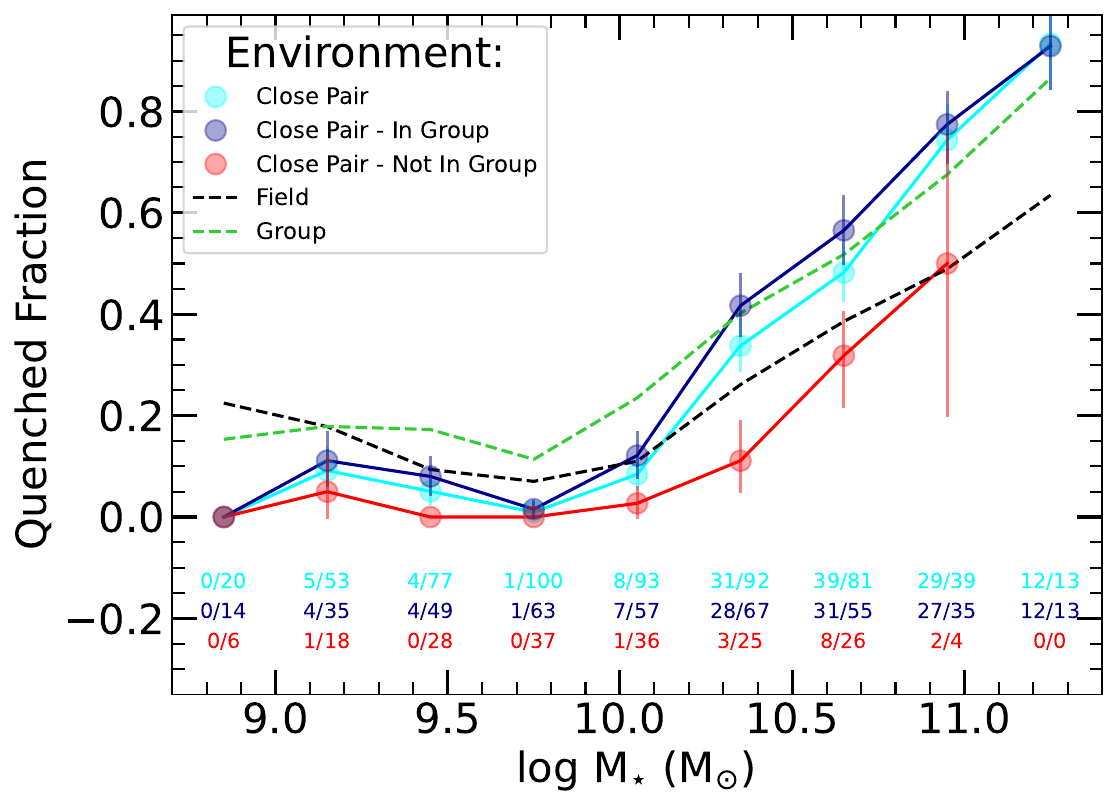}
\caption{The fraction of quenched galaxies per given stellar mass bin for close pairs, splitting close pairs that are also found within groups and those that are not in groups. The quenched fraction represents the ratio of quenched galaxies to the total number of galaxies per mass bin. Each mass bin has a width of 0.3 dex, with errors calculated via bootstrap resampling within each bin. The total number of quenched galaxies and the total number of galaxies for each mass bin are provided below the distributions. The field and group population quenched fraction is depicted with a dash-dot black and green line respectively.}
\label{QF_Pairs_Separate}
\end{figure}

As initially discussed in Section \ref{subsec:Pair Selection}, we include close pairs within groups, this allows some pairs to be classified in both the close pair and group samples. Our results show that SF is statistically suppressed in galaxies within groups, with the largest differences observed at higher stellar masses. Given that approximately two-thirds of the close pairs are located in groups, it raises the question: \textit{are groups driving the results in our close pair sample?} To address this, Figure \ref{QF_Pairs_Separate} revisits Figure \ref{Local_QF} by comparing the overall close pair sample (cyan) with a breakdown of close pairs found in groups (dark blue) and those outside of groups (red). For reference, the quenched fraction distributions of the field and group galaxies are shown by the black and green dashed lines, respectively.

Figure \ref{QF_Pairs_Separate} shows that at $\log M_{\star} < 10 \: M_{\odot}$, group environments do not appear to strongly influence close pairs, as both close pairs within and outside groups exhibit similar trends, which differ notably from the group trend. However, at $\log M_{\star} > 10 \: M_{\odot}$, the results diverge: close pairs within groups align with the group quenched fraction, while those outside groups have significantly lower quenched fractions. In most cases, these isolated close pairs exhibit lower quenched fractions than the field. This suggests that group environments may have a more pronounced impact on the SF properties of close pairs at higher masses, compared to isolated pairs. Nevertheless, due to the limited number of close pairs outside groups, we cannot definitively conclude that there are significant differences between these two populations. Additionally, 73\% (58 of the 79) of galaxies with $\log M_{\star} > 10 \: M_{\odot}$ are located at $z > 0.06$. As the spectroscopic survey completeness declines with increasing redshift, this introduces significant uncertainty regarding whether these pairs are truly isolated or simply missing a companion galaxy below the survey’s completeness threshold.

For these reasons, we do not further subdivide the close pair sample based on group membership. We emphasise the inherent challenges in identifying close pairs within the literature and the need for a more consistent definition. Future large-scale surveys, such as 4HS \citep{ENTaylor23}, will provide the necessary completeness and statistical power to refine this definition and improve the understanding of pair environments.

\end{document}